\documentclass[floats,preprint,aps,tightenlines,showpacs,nofootinbib]{revtex4}
\usepackage{latexsym}
\usepackage[dvips]{graphics}
\usepackage{epsfig}

\newcommand{\beq}{\begin{equation}}
\newcommand{\eeq}{\end{equation}}
\newcommand{\beqa}{\begin{eqnarray}}
\newcommand{\eeqa}{\end{eqnarray}}
\newcommand{\sr}{\sqrt}
\newcommand{\fr}{\frac}
\newcommand{\mn}{\mu \nu}

\newcommand{\vp}{\varphi}

\def\al{\alpha}
\def\lsim{\mathrel{\rlap{\lower4pt\hbox{\hskip1pt$\sim$}}
    \raise1pt\hbox{$<$}}}         %less than or approx. symbol
\def\gsim{\mathrel{\rlap{\lower4pt\hbox{\hskip1pt$\sim$}}
    \raise1pt\hbox{$>$}}}

\def\title#1{\begin{center}{\Large\bf #1}\end{center}}
\def\author#1{\vskip 5mm \begin{center}{#1}\end{center}}
\def\address#1{\begin{center}{\it #1} \end{center} }

\begin{document}

\title{Instanton-inspired Model of QCD Phase Transition and Bubble
Dynamics }
\author{%
Chueng-Ryong Ji$^{a,}$\footnote{E-mail address: crji@unity.ncsu.edu}, %
Gungwon Kang$^{b,}$\footnote{E-mail address: gwkang@kias.re.kr \\
Present address: Dept. of Supercomputing Applications at KISTI
(Korea Institute of Science and Technology Information),
Eoeun-Dong 52, Yuseong-Gu, Daejeon 305-806, Korea }
and %
Jungjai Lee$^{c,}$\footnote{E-mail address: jjlee@daejin.ac.kr} %
}

\address{%
$^a$Department of Physics, North Carolina State University,
Raleigh, NC 27695-8202, USA\\
$^b$School of Physics, Korea Institute for Advanced Study,
207-43 Cheongryangri-Dong, Dongdaemun-Gu, Seoul 130-012, Korea\\
$^c$Department of Physics, Daejin University, GyeongGi, Pocheon
487-711, Korea %
}

\begin{abstract}

We have reinvestigated the collision of gluonic bubbles in a SU(2)
model of QCD which was studied by Johnson, Choi and Kisslinger in
the context of the instanton-inspired model of QCD phase
transition bubbles with plane wave approximation. We discuss
treacherous points of the instanton-inspired model that cause the
violation of causality due to the presence of imaginary gluon
fields. By constructing a new slightly modified Lorentzian model
where we have three independent real gluon fields, we reanalyzed
the process of bubble collisions. Our numerical results show some
indication of forming a bubble wall in colliding region.

\end{abstract}

\pacs{%
}

\maketitle

\newpage

\section{Introduction}

With the current advances of Relativistic Heavy Ion
Collider(RHIC)~\cite{RHICwebsite} and Cosmology in the
Laboratory(COSLAB)~\cite{COSLABwebsite} programs, there is a
growing interest in discussing the self-organizing nature common
to many different physical systems. In particular, the phase
transition and the defect formation are the paramount examples of
the physical phenomena due to the self-organizing nature of
many-body systems. In the RHIC facilities, vigorous analyses are
underway to find if the evidences of forming the quark-gluon
plasma indeed reveal in the experimental data by smashing the
heavy ions. Also, the main emphasis of the COSLAB program is to
exploit the analogies between the phase transitions in the
universe at the time of the hot Big Bang and the observable
transitions in the condensed matter systems at low temperatures.
Consequently, the physical systems experimented in the RHIC and
COSLAB programs are very diverse, spanning hadronic matter,
superfluid Helium, superconductors, liquid crystals, cosmic
strings, etc..  Essential theoretical tools to study these
physical systems are provided by the quantum field theories (QFTs)
including the quantum chromodynamics (QCD), the toy models of QCD
such as Thirring model~\cite{Thirring} and sine-Gordon
model~\cite{sine-Gordon}, and many other dynamical model theories
describing the strongly correlated systems. The description of the
above self-organizing nature with the QFTs, however, cannot rely
on the ordinary perturbative approach because the strong
correlations among the quanta in the formation of new phases are
intrinsically nonperturbative going beyond the realm of the
quantum mechanical von Neumann's theorem~\cite{vonNeumann}.

Consideration of phase transitions in early universe has a rather
long history~\cite{col,hms} and relevant bubble dynamics has been
discussed frequently in the past~\cite{manyworks}. Recently, the
domain walls in QCD have been discussed as examples of topological
defects which may play an important role in the evolution of the
early universe soon after the QCD phase
transition~\cite{Forbes:2000et,ae,cst}. In particular, the
possibility of primordial magnetic field generation has been
discussed as a consequence of the existence of QCD domain
walls~\cite{Forbes:2000et}. Subsequently, a claim for the
formation of cosmological color magnetic wall in the universe was
also made by utilizing the QCD instanton
solutions~\cite{Kisslinger:2002py,Kisslinger:2002mu,Johnson:2003ti}.

Due to the topological configurations of the fields, the domain
walls and the instanton solutions may be stable. Thus, it has been
suggested that the QCD domain walls may be detectable at RHIC
facilities~\cite{Forbes:2000et}. In the COSLAB, reproducing black
hole radiation phenomena ({\it i.e.}, the so-called Hawking
radiation~\cite{Hawking}.) in the table top experiments may be
realized by the phonon radiation from a dumb hole in acoustic
geometry~\cite{dumbhole}. Also discussed is testing various black
hole solutions of the Einstein equation in the
laboratory~\cite{vachaspati}. Similarly, the COSLAB
activities~\cite{COSLABwebsite} may be applicable as well to study
the collisions of the domain walls or the so-called bubble
collisions in the laboratory.

Furthermore, the existence of these topological configurations in
the universe may give an impact on our understanding of the dark
matter which has been one of the biggest puzzles in the cosmology
for more than a decade. Already, the QCD domain wall has been
related to the axion domain wall~\cite{Forbes:2000et}. However,
the more direct connection between the dark matter and the QCD
vacuum condensates has been indicated by the recent molar mass
estimate of dark matter from the strong gravitational lensing in
the galaxy cluster~\cite{MJ}. This estimate also supported by the
analyses of the universal rotation curves in spiral
galaxies~\cite{Persic} revealed that the mass scale of the dark
matter particles may be very close to the QCD scale
$\Lambda_{QCD}$~\cite{MJ,Ariel}. Thus, further detailed studies of
the QCD vacuum properties such as the QCD domain walls and the QCD
instanton solutions~\cite{hooft,instanton} are necessary to
confirm any connection between the QCD vacuum and the dark matter
in the universe.

Motivated by these previous works, we discuss the bubble dynamics
in nonabelian Yang-Mills theory as a prototype of QCD in this
work. In particular, we reanalyze the formation of cosmological
color magnetic wall with and without following the previous
procedure discussed by Kisslinger, et
al.~\cite{Kisslinger:2002py,Kisslinger:2002mu,Johnson:2003ti} and
present both the treacherous points of the previous analysis and
some new progresses made in the present analysis. Although the
potential existence of cosmological magnetic wall and the
importance of its implication on the dark matter and cosmology are
undeniable, our analysis indicates that simplified model
calculations are not capable of verifying its existence and much
more careful studies would be necessary to confirm the
possibility.

The paper is organized as follows. In the next section (Section
II), we briefly go over the equations of motion and the
energy-momentum tensor in the instanton-inspired model of the QCD
phase transition bubbles. In Section III, we solve the derived
equations numerically in a plane-wave model describing the
collision of the two bubbles in the vicinity of the contacting
surface: (a) with the Wick rotation procedure followed by the
previous analysis in Ref.~\cite{Johnson:2003ti}, (b) with our own
method strictly valid in Lorentzian space. In Section III (a), we
reproduce the results obtained by the previous analysis and
present the error estimates of these results to discuss the
treacherous points in the analysis. Moreover, we point out that
the way of solving the system of equations having a constraint
should be corrected from the previous analysis. By solving the
constraint equation first and applying it to the dynamical one, we
show that the numerical results are very different from those in
Ref.~\cite{Johnson:2003ti}. In Section III (b), it is pointed out
that the instanton-inspired model studied in
Refs.~\cite{Kisslinger:2002py,Kisslinger:2002mu,Johnson:2003ti}
should be modified due to the presence of imaginary gluon fields.
By constructing a new slightly modified Lorentzian model, we
present our own results indicating the formation of a bubble wall
in colliding region. Conclusions and Discussions follow in Section
IV. In the appendix, more details of error estimates are
presented.

\section{Instanton-inspired model of QCD phase transition bubbles}

During the time interval between $10^{-5}$-$10^{-4}$s when the
universe passed through the critical temperature $T_{\rm c}\simeq
150$ Mev for the chiral phase transition, the quark-hadron phase
transition (QHPT) occurs. If QHPT occurs as a first order phase
transition, bubbles could form. Recently, a model for describing
bubble collisions has been formulated in
Refs.~\cite{Kisslinger:2002py,Kisslinger:2002mu,Johnson:2003ti}.
Since the main structure of the bubble must be gluonic in nature,
pure gluodynamics was used. In this section, we briefly review the
instanton-inspired model of QCD bubble walls in
Refs.~\cite{Kisslinger:2002py,Kisslinger:2002mu,Johnson:2003ti}.

The action for pure glue is
 \beq
 S = -\frac{1}{4}\int\; d^4x F \cdot F,
 \label{glue}
 \eeq
where
 \beq
\label{G}
 F_{\mu\nu} = \partial_{\mu} A_\nu -  \partial_{\nu}A_\mu
-i g [A_\mu,A_\nu] .
 \eeq
For simplicity SU(2) color gauge field can be considered;
 \begin{eqnarray}
 \label{ymfield}
A_{\mu} & = & A^a_{\mu}(\sigma^a/2)
 \end{eqnarray}
with the Pauli matrix $\sigma_a$ satisfying
$[\sigma^a,\sigma^b]=2i\epsilon^{abc}\sigma^c$ and
$Tr(\sigma^a\sigma^b)=2\delta^{ab}$. Thus, we have
 \begin{eqnarray}
 \label{ymtensor}
F^a_{\mu\nu} = \partial_{\mu} A^a_{\nu} - \partial_{\nu} A^a_{\mu}
+ g\epsilon^{abc}A^b_{\mu} A^c_{\nu},
 \end{eqnarray}
and the field equations in the Lorentz gauge, $\partial^{\mu}
A^a_{\mu}=0$, become
 \begin{eqnarray}
 \label{EOMA}
\partial_\mu\partial^\mu A^a_\nu
+ g\epsilon^{abc}(2 A^b_\mu\partial^\mu A^c_\nu - A^b_\mu
\partial_\nu A^{\mu c}) + g^2\epsilon^{abc}\epsilon^{cef}A^b_\mu
A^{\mu e}A^f_\nu = 0.
 \end{eqnarray}

As is well known, the Euclidean equation of Eq.(\ref{EOMA}) has an
instanton solution given by
 \beq
 \label{inst}
\tilde{A}_\mu^{a,{\rm inst}}(x) = \frac{2}{g}
\tilde{\eta}^{a}_{\mu\nu}\fr{x^\nu}{(x^2 + \rho^2)},
 \eeq
where $\rho$ is the instanton size, the
$\tilde{\eta}^{a}_{\mu\nu}$ symbol is defined in
Ref.~\cite{hooft}, and contractions are defined in Euclidean
metric $\tilde{g}_{\mn}={\rm diag} (1,1,1,1)$ for $\{ x^{\mu} \}
=(x^4,x^1,x^2,x^3)$. Hence, $x^2=(x^4)^2 +\vec{x}^2$. Note that
the gauge field strength for this solution is self-dual. This
instanton connects points in two QCD vacua which differ by one
unit of topological winding number.

Inspired by this instanton solution, Johnson, Choi and
Kisslinger~\cite{Johnson:2003ti} considered a certain class of
Euclidean solutions in the form of
 \beq
\tilde{A}^a_{\mu}(x) = \tilde{\eta}^a_{\mn} \tilde{W}^{\nu}(x).
\label{AW}
 \eeq
By substituting it in the Euclidean action of Eq.(\ref{glue}) and
eliminating all $\tilde{\eta}^a_{\mn}$, they get
 \beqa
\tilde{\mathcal{L}}^{\rm glue}(\tilde{W}) &=&
-\frac{1}{4}\tilde{F}^a_{\mu\nu}\tilde{F}^{\mu\nu a}
\label{ActionEL} \\
&=&-\frac{1}{2}\biggl[ 2(\partial_\mu \tilde{W}_\nu)^2 +
(\partial_\mu \tilde{W}_\mu)^2 + 4 g(\tilde{W}^\mu
\tilde{W}^\nu\partial_\mu \tilde{W}_\nu - \tilde{W}^2\partial_\mu
\tilde{W}^\mu) + 3 g^2 \tilde{W}^4 \biggr]
 \label{ActionEII}
 \eeqa
in Euclidean space $\tilde{g}_{\mn}={\rm diag} (1,1,1,1)$. Then,
by taking the inverse Wick rotation, they obtained a Lorentzian
Lagrangian corresponding to Eq.(\ref{ActionEII}) given by
 \beq
\mathcal{L}^{\rm glue}(W) = -\frac{1}{2}\biggl[ 2(\partial_\mu
W_\nu)^2 + (\partial_\mu W_\mu)^2 + 4 g(W^\mu W^\nu\partial_\mu
W_\nu - W^2\partial_\mu W^\mu) + 3 g^2 W^4 \biggr],
 \label{ActionII}
 \eeq
where $W^{\mu}(x)$ is the analytic continuation of
$\tilde{W}^{\mu}(x)$ associated with Wick rotation $x^4=\tau=it$,
and contractions are defined in Minkowski space $g_{\mn}={\rm
diag} (-1,1,1,1)$. Now the field equations (\ref{EOMA}) can be
expressed in terms of $W^{\mu}(x)$ fields as~\footnote{It should
be pointed out that the direct application of the action principle
to the Lagrangian in Eq.~(\ref{ActionII}) does not reproduce the
field equation (\ref{EOMW}). This follows because, through the
ansatz (\ref{AW}), the number of unknown functions is reduced from
$3 \times 4=12$ for $A^a_{\mu}$ to $4$ for $W^{\mu}$ so that not
all of $A^a_{\mu}$ are independent fields. Thus, it should be
understood that the instanton-inspired model is defined as the
equation (\ref{EOMW}).}
 \begin{eqnarray}
 \label{EOMW}
\partial^2 W_\mu &=& 2 g^2 W^2 W_\mu - 2 g W_\mu(\partial_\al W^\al)
+ 2 g W_\al \partial_\mu W^\al,
 \end{eqnarray}
and the gauge condition becomes
 \begin{eqnarray}
 \label{gcondz}
\epsilon_{\mu\nu\alpha\beta}\partial^\mu W^\nu & = &
\partial_\beta W_\alpha - \partial_\alpha W_\beta .
 \end{eqnarray}
This gauge condition can be expressed as an anti-self-duality
condition for the differential form of the vector field ${\bf W}$,
{\it i.e.}, ${}^*d{\bf W} = - d{\bf W}$. Further restricting the
form of solutions to
 \beq
W^{\mu}(x) = \fr{2}{g} x^{\mu} F(x) ,
 \label{Fform}
 \eeq
one finds
 \beq
 \label{EOMF}
\left( \partial^2 F + 4F x \cdot \partial F + 12 F^2 - 8x \cdot x
F^3 \right) x_{\mu} +2\left( 1-2x \cdot x F \right)
\partial_{\mu}F = 0.
 \eeq
Here $x \cdot x =x^2=-t^2 +\vec{x}^2$. The gauge condition of
Eq.(\ref{gcondz}) becomes
 \beq
 \label{gcondzF}
\epsilon_{\mu\nu\alpha\beta}x^\nu \partial^\mu F = x_\alpha
\partial_\beta F - x_\beta \partial_\alpha F .
 \eeq
Since Eq.~(\ref{EOMF}) implies $\partial_{\mu}F \sim x_{\mu}$, any
solution $F$ of Eq.~(\ref{EOMF}) satisfies the gauge condition
above automatically. Multiplying $x^{\mu}$ to Eq.~(\ref{EOMF}),
one obtains
 \beq
 \label{EOMF2}
\partial^2 F + \frac{2}{x^2}x \cdot \partial F + 12 F^2
-8 x \cdot x F^3 = 0.
 \eeq
By substituting it into Eq.~(\ref{EOMF}) again, we have
 \beq
\left( \partial_{\mu}F -\fr{x \cdot \partial F}{x^2} x_{\mu}
\right) \left( 1-2x \cdot x F \right) =0.
 \eeq
Thus, one finds that
 \beq
\partial_{\mu}F = \fr{x\cdot \partial F}{x\cdot x} x_{\mu}, \qquad
{\rm or} \qquad F(x) = \fr{1}{2x\cdot x}=\fr{1/2}{-t^2+\vec{x}^2} .
 \eeq
For the first case, Eq.~(\ref{EOMF}) becomes equivalent to
Eq.~(\ref{EOMF2}). For the second case, on the other hand,
Eq.~(\ref{EOMF}) becomes
 \beq
\partial^2 F + 4F x \cdot \partial F + 12 F^2 - 8x \cdot x
F^3 =0 ,
 \eeq
which is satisfied by $F=(2x\cdot x)^{-1}$. Notice that this
equation differs from Eq.~(\ref{EOMF2}) only by $4x\cdot \partial
F (F-1/(2x\cdot x))$ which vanishes in this case. It should be
pointed out that, in contrast to Ref.~\cite{Johnson:2003ti}, here
we have not assumed that $F$ is a function of $x^2=x\cdot x$ only
although Eq.~(\ref{EOMF}) seems to indicate it. One can also see
that there is another exact solution of Eq.(\ref{EOMF2}) given by
 \beq
F(x^2) = \fr{1}{-t^2+\vec{x}^2 +\rho^2}.
 \label{Instanton}
 \eeq
In Ref.~\cite{Johnson:2003ti} this solution was considered as an
analytic continuation of the Euclidean instanton solution in
Eq.(\ref{inst}) starting at $t=0$.

From $\mathcal{L}^{\rm glue}(W)$ in Eq.~(\ref{ActionII}), we get
the following energy-momentum tensor given by
 \begin{eqnarray}
 \label{EMT}
T_{\mu\nu} &=& \sum_a( g^{\rho \sigma} F^a_{\mu\rho}
F^a_{\nu\sigma} -\frac{1}{4}g_{\mu \nu} F^{a \rho\sigma}
F^a_{\rho\sigma})
\nonumber\\
&=& 2\partial_{(\mu} W_{\nu)} \partial^\rho W_\rho +g_{\mu
\nu}\partial^\rho W^\sigma \partial_\rho W_\sigma -2g \biggl[W^2
\partial_{(\mu} W_{\nu)} -\left( W_\mu W_\nu -g_{\mu \nu}W^2 \right)
\partial^\rho W_\rho  \nonumber\\
&& -g_{\mu \nu}W^\rho W^\sigma \partial_\rho W_\sigma \biggr]
+2g^2 \left( g_{\mu \nu}W^4 - W_\mu W_\nu W^2 \right) +g_{\mu\nu}
{\cal L}^{\rm glue}
\nonumber\\
&=& \frac{4}{g^2}\biggl[ 8 F x_{(\mu} \partial_{\nu)} F +2x \cdot x
\;
\partial_\mu F \partial_\nu F+ g_{\mu \nu}[12F^2 +4F x \cdot
\partial F + x \cdot x \; \partial^\rho F \partial_\rho F ] \nonumber\\
&& -8 F^3 \left( x \cdot x \; F -2 \right) \left( x_\mu x_\nu -x
\cdot x \; g_{\mu\nu} \right) +4\left( x_\mu x_\nu x \cdot \partial
F -x\cdot x x_{(\nu}\partial_{\mu)}F \right) F^2 \biggr] \nonumber
\\
&&   +g_{\mu\nu}{\cal L}^{\rm glue}.
\end{eqnarray}
Note that the energy-momentum tensor used in
Ref.~\cite{Johnson:2003ti} has wrong sign for the second term on
the first line above. If we insert the exact solution $F$ in
Eq.~(\ref{Instanton}) into the energy-momentum tensor
Eq.~(\ref{EMT}), $T_{\mu \nu}$ appears to be zero. This fact can
be expected from the self dual property of the instanton solution
for the gauge field strength in this system. In general, it is
well-known that topological gauge fields such as self or anti-self
dual solution provide the vanishing energy density. This is
contrary to the results of non-vanishing energy density presented
in Refs.~\cite{Kisslinger:2002py,Kisslinger:2002mu}.

Nevertheless, it should be pointed out that $F$ is singular at
points where $\vec{x}^2 - t^2 + \rho^2 =0$ while the corresponding
Euclidean solution is regular in whole Euclidean space.
Consequently, $T_{\mu \nu}$ is not zero everywhere in Minkowski
space, but singular at those divergent points of $F$. This
singular behavior is analogous to what occurs in type II
superconductivity~\cite{Shuryak}. Although the magnetic field can
not penetrate the superconductor due to the Meissner effect, it is
well known that the strong magnetic field can destroy the cooper
pairs and generate a pinhole in the type II superconductor. In the
ideal superconductor this appears as a singular solution for the
magnetic field. However, in realistic superconductor this
singularity melts due to the interaction with environmental
substances. Thus the singular energy momentum tensor in Minkowski
space might be interpreted to generate a strong color magnetic
wall in the universe. This sort of interpretation may support the
previous discussions on the QCD domain walls and the formation of
cosmological color magnetic walls in
Refs.~\cite{Kisslinger:2002mu,Forbes:2000et}.

Motivated by this, we would like to discuss the possibility of the
formation of this strong color magnetic wall in the process of
bubble collisions in gluonic dynamics. In the next section, we
consider two models of bubble collisions and analyze the time
evolution of field configuration depending on the initial
conditions.

\section{Bubble Collisions in plane wave models}

In general, the analysis of bubble collision is not an easy task,
but involves a lot of numerical computations. The authors in
Ref.~\cite{Johnson:2003ti} considered a $(1+1)$ dimensional
model\footnote{This terminology might be misleading since this
model is actually defined only in ($3+1$) dimensions due to the
inappropriateness of $\tilde{\eta}$-symbols in ($1+1$)
dimensions.} in order to see whether an interior gluonic wall is
formed during the collision of two QCD bubbles. This model may be
appropriate to describe the collision process in the vicinity of
contact surface, especially, in the case that the size of
colliding bubbles is very large. In such situations the bubble
walls around a colliding surface can be treated as plane waves.
After we briefly summarize the results of
Ref.~\cite{Johnson:2003ti}, we present our new analysis.

\subsection{Instanton-inspired model}

\begin{figure}[tbp]
\begin{minipage}[b]{.46\linewidth}
\centering
\resizebox{7.8cm}{7.8cm}{\includegraphics{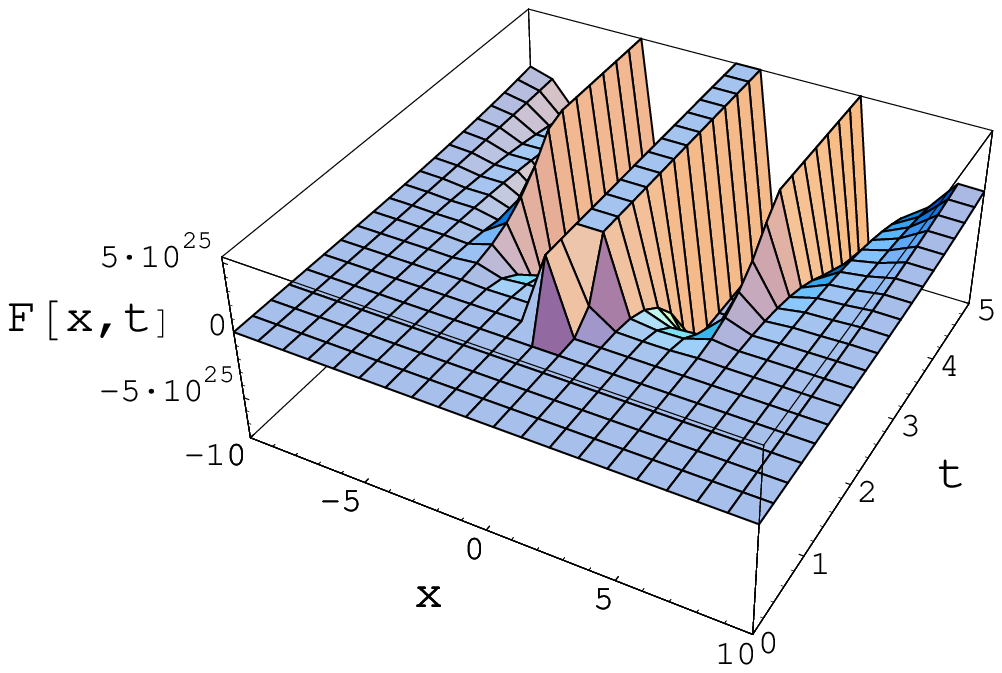}}
\caption{The evolution of the field $F(x,t)$ in the collision of
two bubbles. A gluonic wall seems to form in the middle.}
 \label{figJCK}
\end{minipage}\hfill
\begin{minipage}[b]{.46\linewidth}
\centering
\resizebox{7cm}{7cm}{\includegraphics{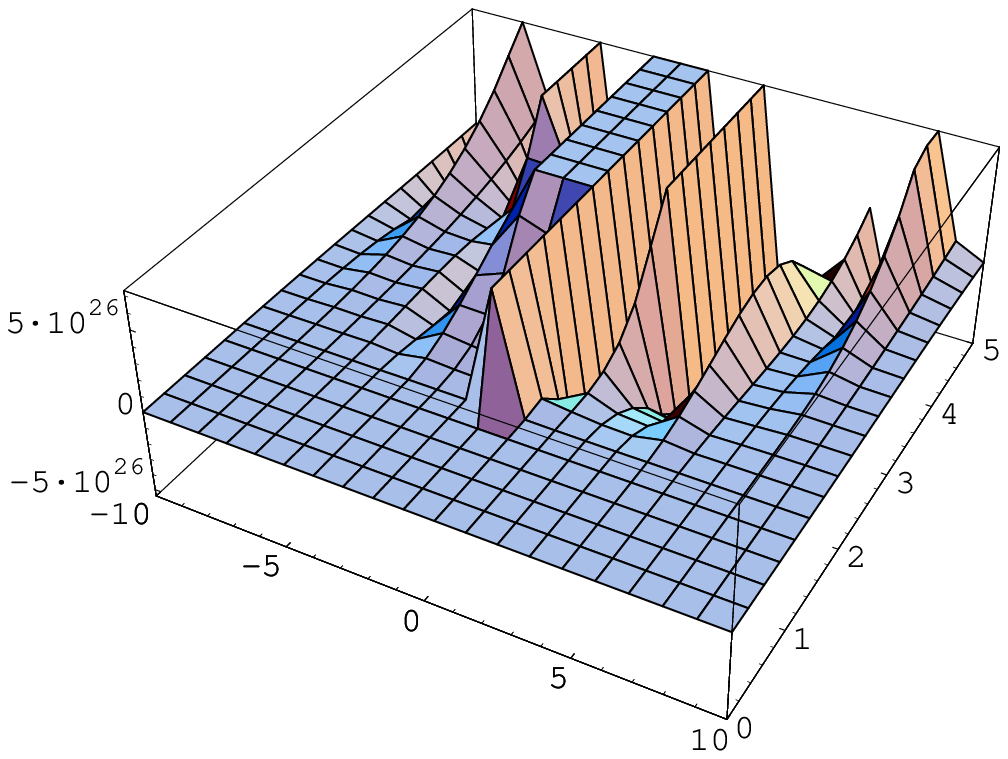}}
 \put(-129,24){$x$}
 \put(-15,58){$t$}
 \put(-202,132){$C(x,t)$}
\caption{Numerical error estimated by the gauge constraint equation,
$C(x,t)= x \partial_t F +t \partial_x F =0$.}
 \label{fig2Error}
\end{minipage}
\end{figure}

Suppose the collision occurs in the $x$-direction. Then the gauge
fields in the $(1+1)$ dimensional model~\cite{Johnson:2003ti}
would be independent of $y, z$-coordinates in the vicinity of
contacting surface. Taking $W_2=W_3=0$ in $W_{\mu}$ in
Eq.(\ref{EOMW}), the authors in Ref.~\cite{Johnson:2003ti}
obtained the following equations for $W_0(x,t)$ and $W_1(x,t)$;
 \begin{eqnarray}
 \label{eom1}
\partial^2 W_0 &=& 2 g^2 (W_1^2 -W_0^2) W_0 -2 g W_0\partial_x W_1
+ 2 g W_1 \partial_t W_1,  \nonumber\\
\partial^2 W_1 &=& 2 g^2 (W_1^2 -W_0^2) W_1 + 2 g W_1\partial_t W_0
- 2 g W_0 \partial_x W_0
 \end{eqnarray}
with the gauge condition
 \begin{eqnarray}
 \label{gcondx}
\partial_x W_0 & = & \partial_t W_1.
 \end{eqnarray}
In this model, the authors of Ref.~\cite{Johnson:2003ti} stated
that Eq.(\ref{EOMF2}) becomes~\footnote{Although we follow their
equation to check their numerical results, we found that rewriting
Eq.~(\ref{eom1}) in terms of $F(x)$ gives $-4F^2$ instead of
$-12F^2$ in Eq.~(\ref{PlaneF}). We point out that
Eq.~(\ref{EOMF2}) is valid only in four dimensions.}
 \beq
\partial^2 F = -\fr{2}{x} \partial_x F -12 F^2 +8(-t^2+x^2) F^3
 \label{PlaneF}
 \eeq
with the gauge constraint
 \beq
x \partial_t F +t \partial_x F =0.
 \label{Errorfn}
 \eeq
\\
With the initial data for the two bubble walls at $t=0$ given by
 \beq
F(x,t)|_{t=0} = \fr{1}{(x-3)^2 +\rho^2} +\fr{1}{(x+3)^2 +\rho^2},
\qquad
\partial_t F|_{t=0} =0
 \label{InitialDF}
 \eeq
and a periodic boundary condition
 \beq
F(-10,t) = F(10,t),
 \eeq
the authors in Ref.~\cite{Johnson:2003ti} solved Eq.~(\ref{PlaneF})
numerically. Since the distance of two peaks is $6$, we may expect
that the time of collision would be $t \sim 6$ if each peak moves
with the speed of light. However, due to the smooth profile of this
initial data, the actual collision would occur much before this time
scale. We have reproduced all of their published results. As an
example, the function $F(x,t)$ is shown in Fig.~\ref{figJCK} which
corresponds to their Fig.~2 in Ref.~\cite{Johnson:2003ti}. Based on
this numerical result, they stated that they did find a gluonic
structure evolving at the colliding region although they mentioned
the accuracy of the calculation for $t > 1.0$ is limited.

However, we found that this numerical result is not really
reliable for several reasons. First of all, notice that the
numerical value of $F(x,t)$ drastically changes from the order of
$1$ to $\sim 10^{26}$ around $t \simeq 1.1$. It indicates that
some singularities occur around at $t \simeq 1.1$ so that the
accuracy of the numerical calculations is severely limited beyond
this point. Indeed we find that the numerical error estimated by
the gauge constraint equation (\ref{Errorfn}) rapidly grows as
shown in Fig.~\ref{fig2Error}. While the gauge condition consists
of the two terms, the combination of the two terms should not be
larger than each individual term in order for the error to be
reasonably small. However, Fig.~\ref{fig2Error} shows unreasonably
large error beyond $t\simeq 1.1$. The authors in
Ref.~\cite{Johnson:2003ti} have also solved Eq.~(\ref{eom1})
numerically for two fields $W_0(x,t)$ and $W_1(x,t)$, and reached
the same conclusion of the formation of gluonic walls. Even in
this case, however, we found severe numerical errors as explained
in the appendix.

Secondly, note that the singular behavior colliding at $t\simeq
1.1$ occurs much before the expected collision time scale $t\sim
6$. As we mentioned earlier this may not be unreasonable due to
the smooth profile of the initial data, but in order to clarify
whether this indicates a violation of causality or not it is
necessary to analyze the evolution of the profile in much more
detail.

Finally, we point out that the way of solving the set of coupled
equations (\ref{PlaneF}) and (\ref{Errorfn}) as above should be
corrected. In Ref.~\cite{Johnson:2003ti} $F(x,t)$ is assumed to be
a function of $\xi = x^2-t^2$ only. Consequently, the gauge
constraint equation (\ref{Errorfn}) is satisfied automatically.
Now one may think that the remaining thing to do is solving the
second order partial differential equation (\ref{PlaneF}) for any
given initial value of $F$ with vanishing $\partial_tF|_{t=0}=0$
as in Eq.~(\ref{InitialDF}). Regardless of solving
Eq.~(\ref{PlaneF}), however, the time evolution of $F(x,t)$ is
completely determined by its initial value since $F(x,t)$ is
assumed to be a function of $\xi=x^2-t^2$ only. That is, if
$F=F_0$ at $(x,t)=(x_0,0)$, it should be that, for any position
$(x,t)$ satisfying $x^2-t^2 =\xi_0=x_0^2$ in the future,
$F(x,t)=F(x^2-t^2)=F(\xi_0)=F_0$. In other words, the initial
value of $F$ must evolve without any change along the
$x^2-t^2=\xi= {\rm constant}$ hyperbola. However, we cannot
observe such behavior in Fig.~\ref{figJCK}.

Although $F(x,t)$ is assumed to be a function of $\xi$ only in
Ref.~\cite{Johnson:2003ti}, we show that this assumption is indeed
the most general solution of the gauge constraint equation
(\ref{Errorfn}). Consider the coordinate transformation such that
$\xi=x^2-t^2$ and $\eta=tx$. Then Eq.~(\ref{Errorfn}) becomes
 \beq
\partial_{\eta}F(\xi,\eta) =0,
 \eeq
and so the constraint equation can be solved easily, giving that
$F(x,t)$ is a function of $\xi=x^2-t^2$ only. Consequently,
Eq.~(\ref{PlaneF}) can be rewritten as
 \beq
\xi F''(\xi) + 2F'(\xi) +3F^2(\xi) -2\xi F^3(\xi) =0,
 \label{EOMFxi}
 \eeq
where $F'(\xi)=dF(\xi)/d\xi$. This is simply a second order
ordinary differential equation with respect to $\xi$. In other
words, the system of problem which was expressed in the
two-dimensional space as in Eq.~(\ref{PlaneF}) is actually in the
one-dimensional space described by $\xi : -\infty \sim \infty$,
due to the presence of the gauge constraint equation
(\ref{Errorfn}).
\begin{figure}[tbp]
\begin{minipage}[b]{.46\linewidth}
\centering \resizebox{6cm}{6cm}{\includegraphics{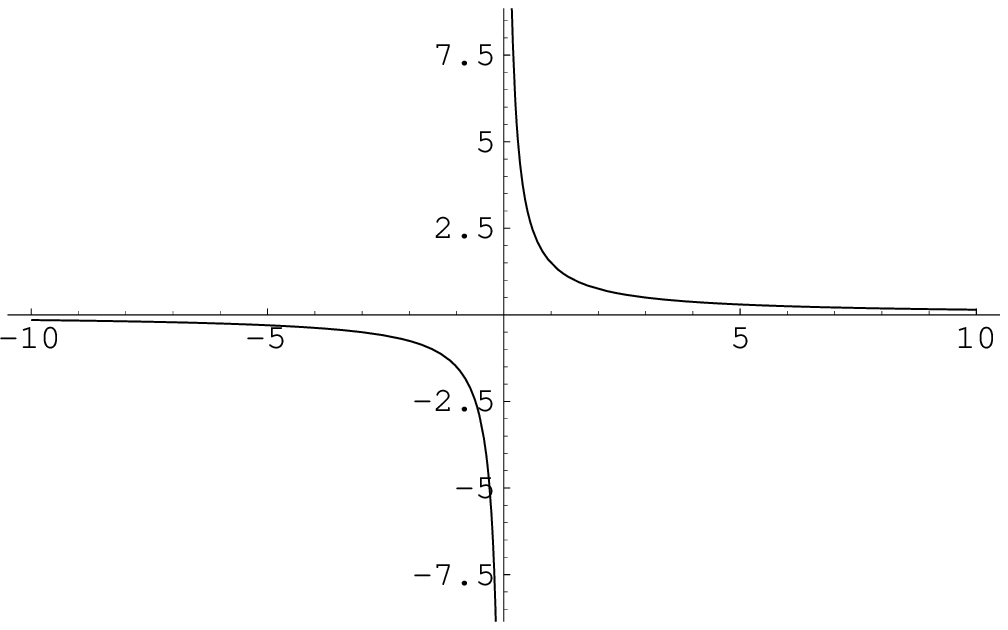}}
 \put(5,81){$\xi$}
 \put(-90,175){$F(\xi)$}
 \caption{  The exact solution Eq.~(\ref{FExact})
in $\xi$-space.}
 \label{FigFExact}
\end{minipage}\hfill
\begin{minipage}[b]{.46\linewidth}
\centering \resizebox{6cm}{6cm} {
\includegraphics{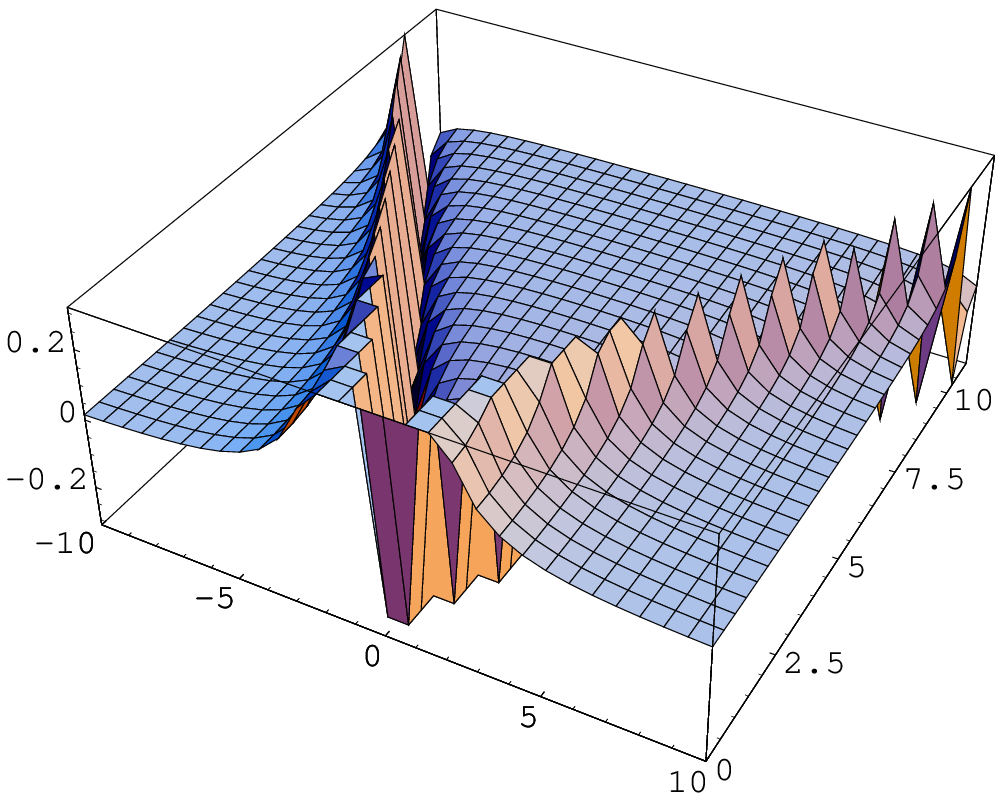}}
 \put(-118,17){$x$}
 \put(-11,48){$t$}
 \put(-170,115){$F(x,t)$}
\caption{The exact solution Eq.~(\ref{FExact}) in $(x,t)$-space.}
 \label{FigFExact3D}
\end{minipage}
\end{figure}
Therefore, giving the initial data for $F(x,t)$ at $t=0$ is
equivalent to assigning the value of $F(\xi)$ for $\xi > 0$. In
the case of Eq.~(\ref{InitialDF}), we have
 \beq
F(\xi) = \fr{1}{(\sr{\xi} -3)^2 +\rho^2} +\fr{1}{(\sr{\xi} +3)^2
+\rho^2}
 \eeq
for $\xi >0$. Now one can easily check that this function above is
not a solution of Eq.~(\ref{EOMFxi}). It implies that one cannot
take an arbitrary function of $F(x,0)$ as an initial data set at
$t=0$ in Eq.~(\ref{PlaneF}). The solution space of
Eq.~(\ref{PlaneF}) alone is much larger than that of
Eq.~(\ref{EOMFxi}). Only some of them corresponding to a specific
class of initial data satisfy Eq.~(\ref{Errorfn}) as well at
arbitrary time.

\begin{figure}[b]
\begin{minipage}[b]{.46\linewidth}
\centering \resizebox{6cm}{5cm}
 {\includegraphics{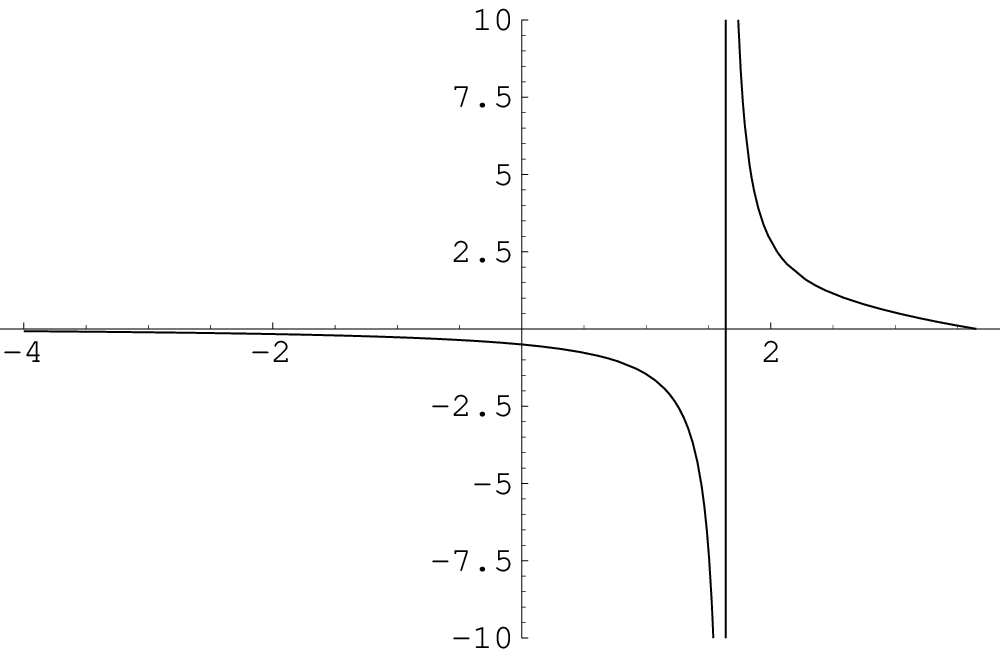}}
 \put(5,68){$\xi$}
 \put(-90,147){$F(\xi)$}
 \begin{flushleft}
(a) $F=-1/2$ and $F'=-3/8$ at $\xi=0$. A singular peak occurs at
$\xi \simeq 1.6$.
\end{flushleft}
\end{minipage}\hfill
\begin{minipage}[b]{.46\linewidth}
\centering \resizebox{6cm}{5cm}
 {\includegraphics{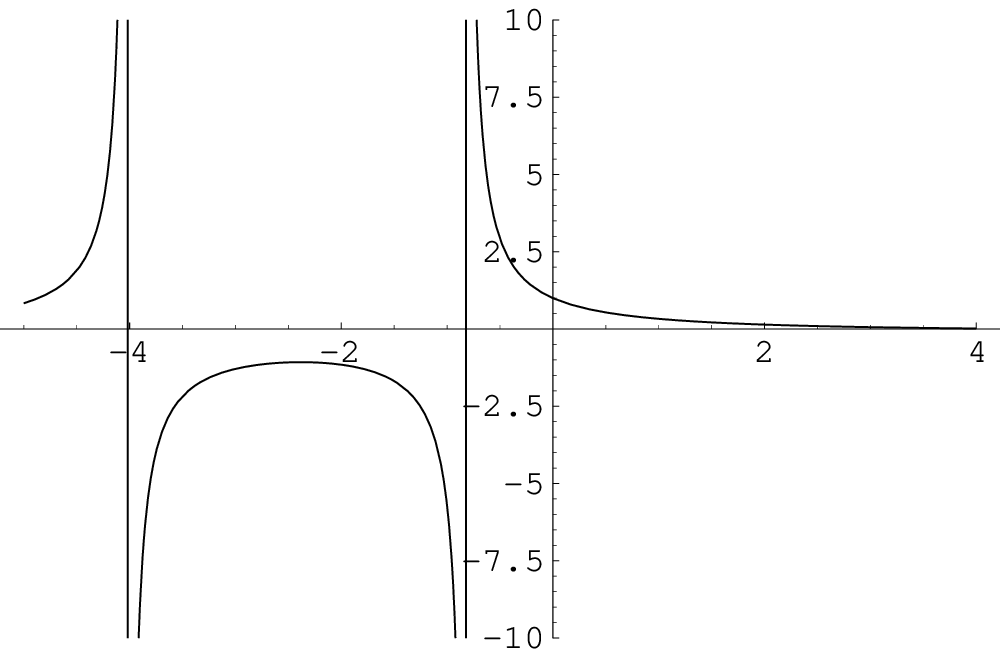}}
 \put(5,68){$\xi$}
 \put(-83,147){$F(\xi)$}
\begin{flushleft}
(b) $F=1$ and $F'=-3/2$ at $\xi=0$. Singular peaks occur at $\xi
\simeq -0.8$ and $-4.0$.
\end{flushleft}
\end{minipage}
\caption{\label{FigFNumerical} Some numerical solutions for
Eq.~(\ref{EOMFxi})}
\end{figure}

We have seen that the correct way of solving the coupled equations
(\ref{PlaneF}) and (\ref{Errorfn}) is to solve Eq.~(\ref{EOMFxi}).
What would be the typical behavior of the solutions for
Eq.~(\ref{EOMFxi})? One can easily see that
 \beq
F(\xi)=\fr{3/2}{\xi}=\fr{3/2}{x^2-t^2}
 \label{FExact}
 \eeq
is an exact solution of Eq.~(\ref{EOMFxi}). We see that the
singular surface propagates along the light cone, {\it i.e.},
$\xi=0$ or $x=\pm t$ as shown in Fig.~\ref{FigFExact3D}. When
$F(\xi)$ is not singular at $\xi=0$, we may have the asymptotic
solution near $\xi=0$ given by
 \beq
 F(\xi) = b -\fr{3b^2}{2} \xi +\fr{11b^3}{6} \xi^2 + \cdots .
 \eeq
The numerical solutions for the cases of $b=1, -1/2$ are shown in
Fig.~\ref{FigFNumerical}, respectively.
\begin{figure}[btp]
\begin{minipage}[b]{.23\linewidth}
\centering \resizebox{4cm}{4cm}
 {
\includegraphics{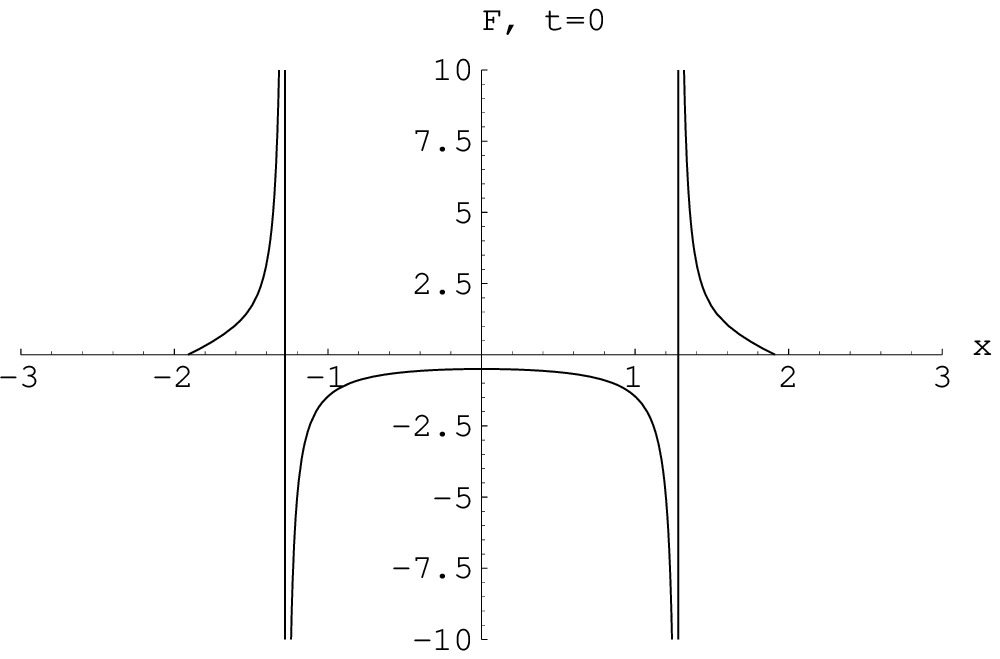}
 }
\end{minipage}
\hfill
\begin{minipage}[b]{.23\linewidth}
\centering \resizebox{4cm}{4cm}
 {
\includegraphics{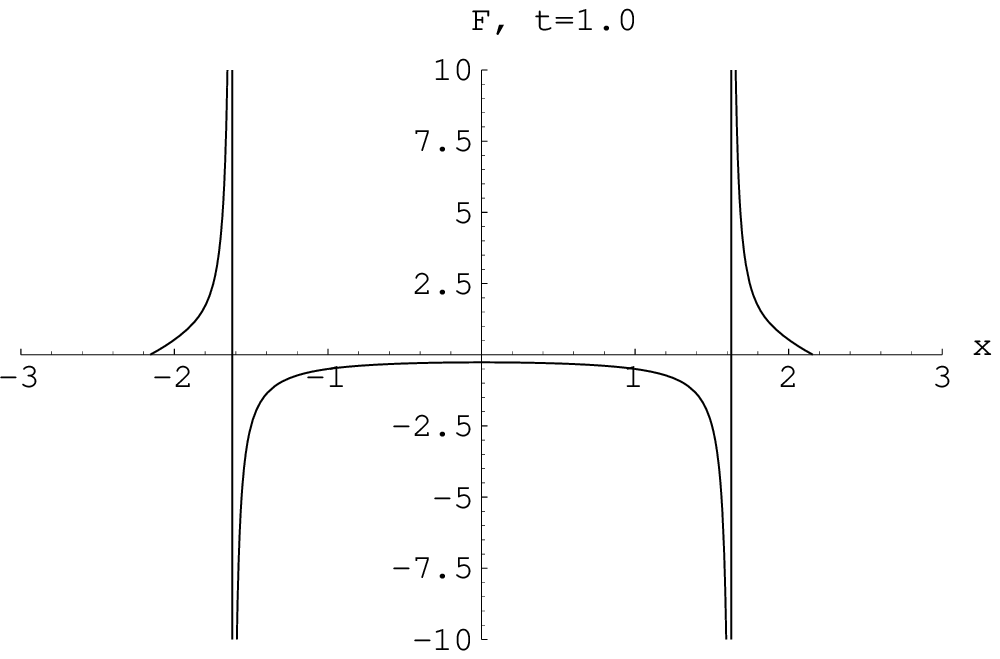}
 }
\end{minipage}
\hfill
\begin{minipage}[b]{.23\linewidth}
\centering \resizebox{4cm}{4cm}
 {
\includegraphics{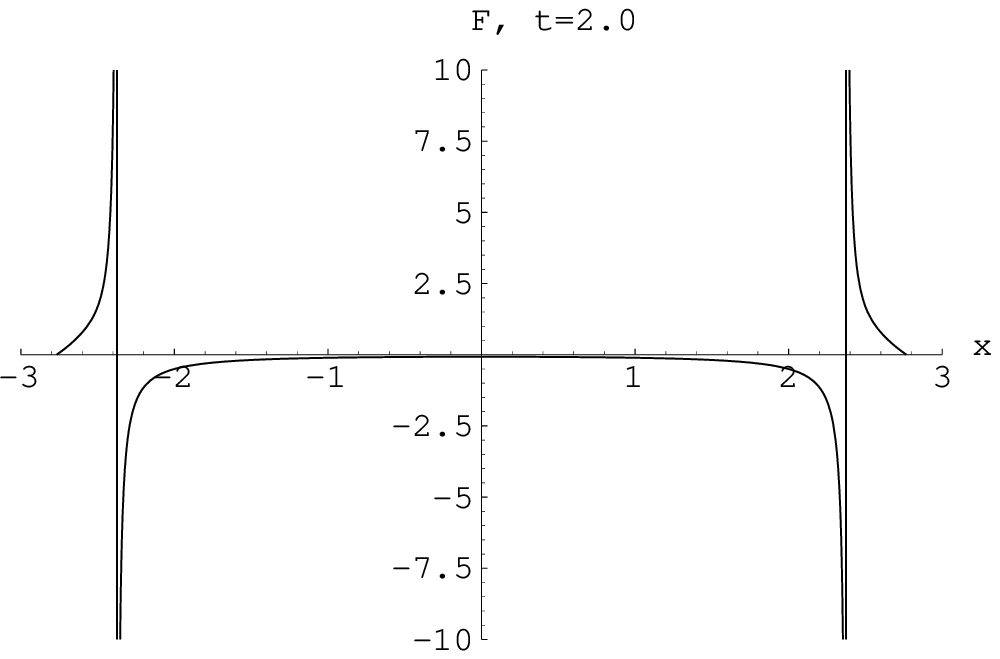}
 }
\end{minipage}
 \hfill
\begin{minipage}[b]{.23\linewidth}
\centering \resizebox{4cm}{4cm}
 {
\includegraphics{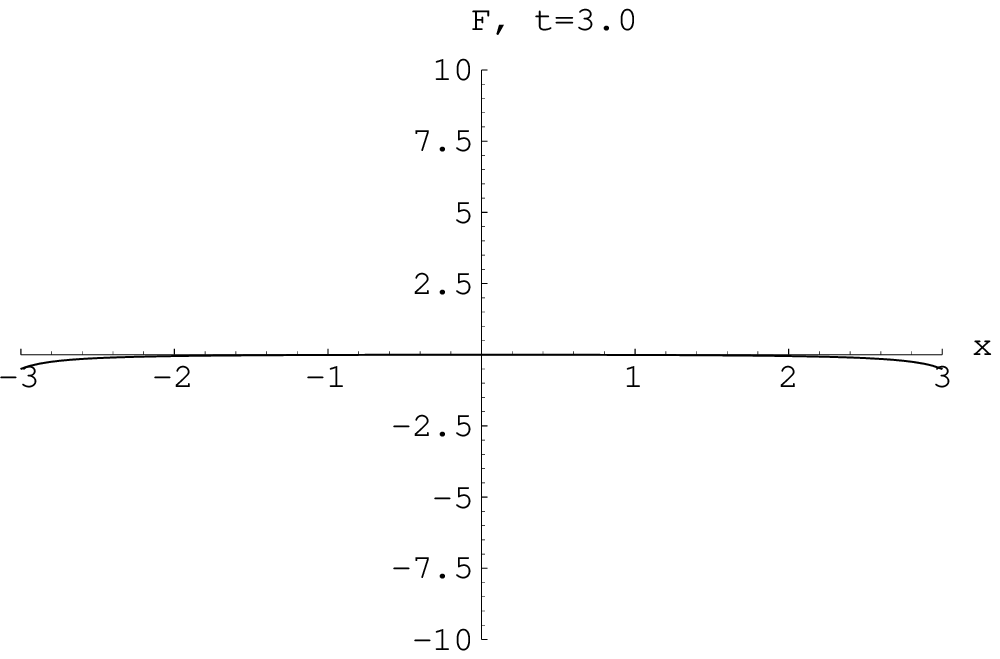}
 }
\end{minipage}
\caption{The time evolution of the numerical solution in
Fig.~\ref{FigFNumerical} (a) is captured at $t=0,\ 1.0,\ 2.0,\
{\rm and} \ 3.0$, respectively.}
 \label{FigFNumericalttaa}
\end{figure}
\begin{figure}[tbp]
\begin{minipage}[b]{.17\linewidth}
\centering \resizebox{3.2cm}{4cm}
 {
\includegraphics{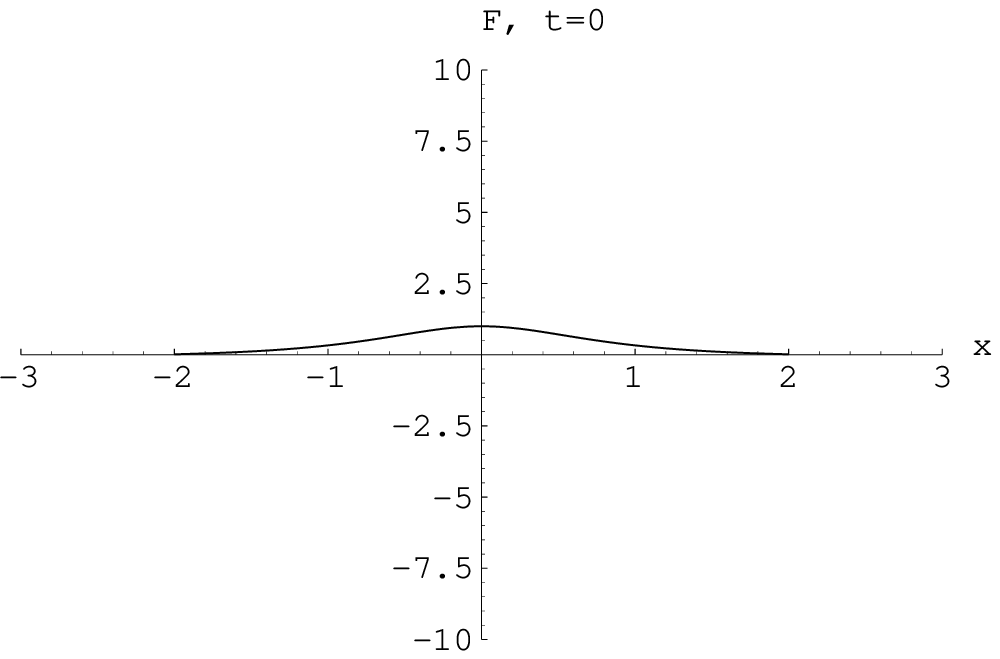}
 }
\end{minipage}\hfill
\begin{minipage}[b]{.17\linewidth}
\centering \resizebox{3.2cm}{4cm}
 {
\includegraphics{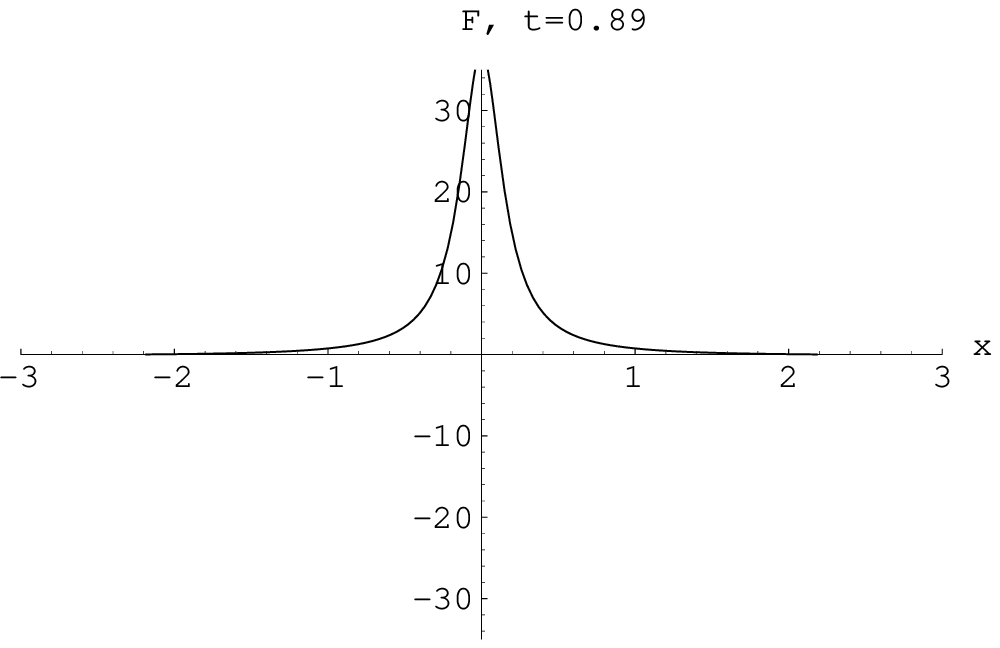}
 }
\end{minipage}
\hfill
\begin{minipage}[b]{.17\linewidth}
\centering \resizebox{3.2cm}{4cm}
 {
\includegraphics{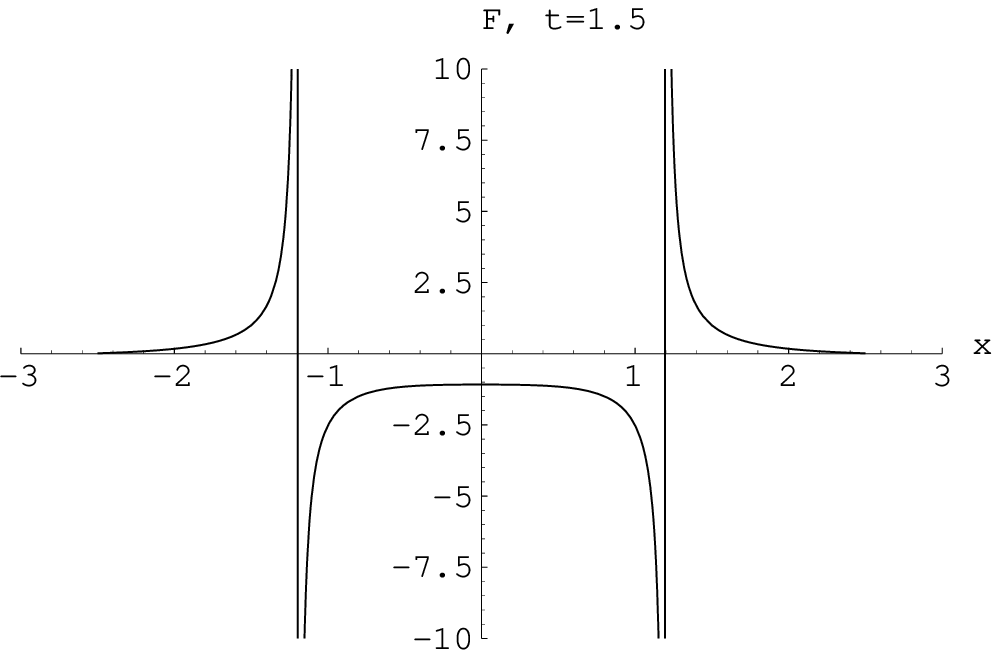}
 }
\end{minipage}
\hfill
\begin{minipage}[b]{.17\linewidth}
\centering \resizebox{3.2cm}{4cm}
 {
\includegraphics{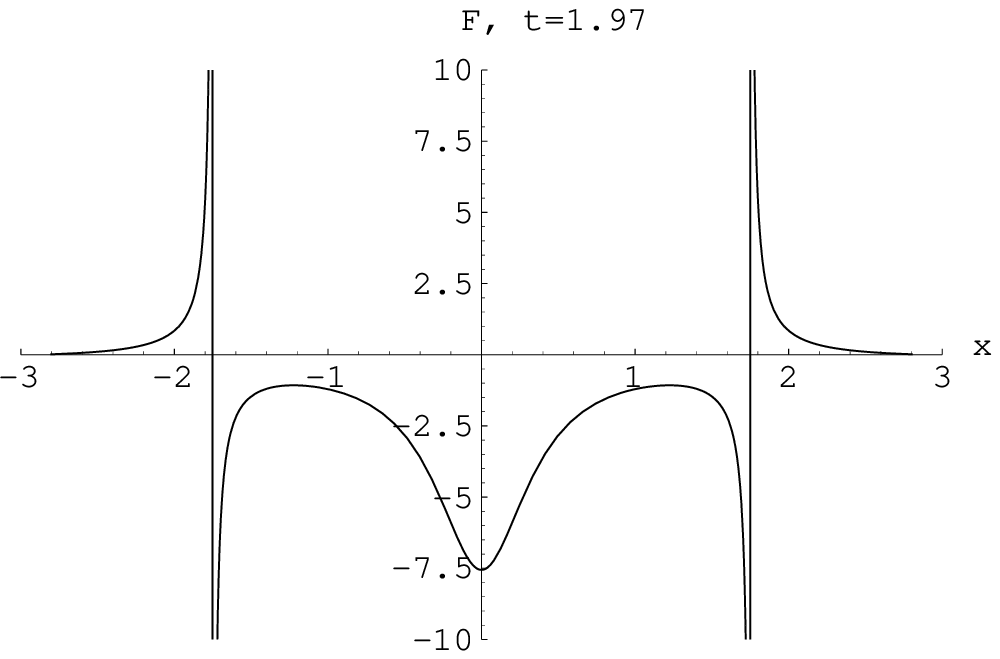}
 }
\end{minipage}
\hfill
\begin{minipage}[b]{.17\linewidth}
\centering \resizebox{3.2cm}{4cm}
 {
\includegraphics{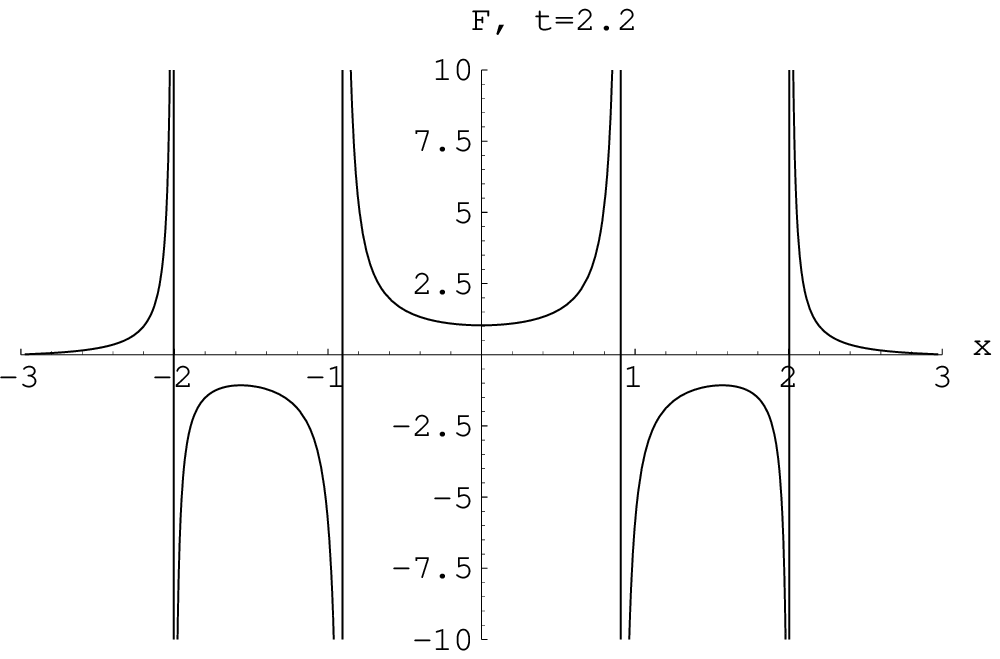}
 }
\end{minipage}
\caption{The time evolution of the numerical solution in
Fig.~\ref{FigFNumerical} (b) is captured at $t=0,\ 0.89,\ 1.5,\
1.97,\ {\rm and} \ 2.2$, respectively.}
 \label{FigFNumericaltt}
\end{figure}
In $(x,t)$-space, Fig.~\ref{FigFNumerical} (a) illustrates a wave
initially having two singular peaks at $x \simeq \pm \sr{1.6}= \pm
1.26$. As shown in Fig.~\ref{FigFNumericalttaa}, these two peaks
simply propagate away from each other along the hyperbolas $x^2-t^2
\simeq 1.6$, and no formation of a wall can be seen in its future
evolution since $|F(x,t)|$ monotonically decreases as $\xi
\rightarrow -\infty$. For the case of Fig.~\ref{FigFNumerical} (b),
the initial wave has a single regular peak located at $x=0$. This
peak quickly grows till $t \simeq \sr{0.8}=0.894$, splits into two
singular peaks which propagate away along the hyperbolas $x^2-t^2
=\xi \simeq -0.8$. At around $t \simeq 2$ a singular peak is formed
again, and splits into two singular peaks which propagate away from
each other along the hyperbolas $x^2-t^2 =\xi \simeq -4$ as shown in
Fig.~\ref{FigFNumericaltt} explicitly.

Therefore, we do not observe the formation of a wall through the
collision of two bubbles in this model. Even if a wall is formed
`spontaneously', as in Fig.~\ref{FigFNumericaltt}, it splits into
two parts immediately, propagating away from each other faster
than the speed of light. The peak does not last in the region of
collision. This property and the violation of the causality should
be expected in the region of $\xi < 0$ basically because $F(x,t)$
is the function of $x^2-t^2=\xi$. In the next section, we discuss
where the problematic features of this instanton-inspired model
come from, and present our new model based on the conventional
Lorentzian theory.

\subsection{New Lorentzian model}

In this section we point out that the instanton-inspired model in
which the dynamics of bubble collisions were studied above is
actually problematic as a Lorentzian theory. Let us go back to the
construction of such model in Eq.~(\ref{EOMW}). Under the Wick
rotation such as $x^4=\tau \rightarrow \tau=it$ and $x^i
\rightarrow x^i$ for $i=1,2,3$ the relationships between fields in
Lorentzian and Euclidean spaces become
 \beq
A^a_t = i \tilde{A}^a_{\tau}, \qquad A^a_i = \tilde{A}^a_{i},
\qquad \eta^1_{t1} =\eta^2_{t2} =\eta^3_{t3} =\tilde{\eta}^1_{\tau
1} = \cdots = -i , \qquad \eta^1_{23} =\tilde{\eta}^1_{23} = 1,
 \eeq
where $\eta^a_{\mn}$ is the $\eta$-symbol in Lorentzian space
corresponding to the Euclidean t'Hooft symbol
$\tilde{\eta}^a_{\mn}$~\cite{Mueller:1991fa}. Thus, the
relationships between colored vector fields $A^a_{\mu}$ in the
original Lorentzian action in Eq.~(\ref{glue}) and fields
$W_{\mu}$ in the instanton-inspired model in Eq.~(\ref{EOMW}) are
indeed
 \beq
A^a_{\mu} = \left( \begin{array}{cccc}
 -iW_1 & -iW_t & W_3 & -W_2   \\
 -iW_2 & -W_3 & -iW_t & W_1  \\
 -iW_3 & W_2 & -W_1 & -iW_t
 \end{array}
 \right) .
 \label{RelationshipLE}
 \eeq
Since the fields $W_{\mu}$ are supposed to be real functions in
the instanton-inspired model, it implies that the dynamics for the
fields $W_{\mu}$ in this theory is not usual as in ghost-like
dynamics in a scalar field system where the scalar field is
replaced by an imaginary field, {\it e.g.}, $-\partial_{\mu}\phi
\partial^{\mu}\phi = \partial_{\mu}\psi \partial^{\mu}\psi$ with
$\phi=i\psi$. Therefore, although the instanton-inspired model in
Eq.~(\ref{EOMW}) was constructed in order to describe bubble
collisions in Minkowski space starting from instantons at $t=0$,
its dynamics is not the correct Lorentzian one due to the
appearance of the imaginary electromagnetic fields. Furthermore,
it gives rise to the indefiniteness in the energy density for the
fields, {\it e.g.}, $T_{00} \sim - (\vec{E}^a)^2 +(\vec{B}^a)^2$.
Notice also that the singular surface of the solution
Eq.~(\ref{Instanton}) in the instanton-inspired model expands
faster than the speed of light. Thus, it appears to violate the
causality.

One may try to redefine some of $W_{\mu}$ fields to absorb the
imaginary numbers. However, such prescription turns out to be
impossible as can be immediately seen in Eq.~(\ref{RelationshipLE}).
Thus, we consider the following model that may describe the
Lorentzian bubble dynamics correctly in the vicinity of contacting
surface;
 \beq
 A^a_{\mu}(x,t) = \left( \begin{array}{cccc}
\psi(x,t) & \vp(x,t) & 0 & 0  \\
0 & 0 & \vp(x,t) & \chi(x,t)  \\
0 & 0 & -\chi(x,t) & \vp(x,t)
 \end{array}
 \right).
 \label{CKLmodel}
 \eeq
This form of fields was obtained from Eq.~(\ref{RelationshipLE}) by
setting $W_2=W_3=0$ with plane wave symmetry along $y$ and $z$
directions.

With this ansatz Eq.(\ref{EOMA}) becomes
 \beqa
\mbox{} && \partial^2 \vp -2g(\chi \vp' -\vp \chi') -2g^2\vp
(\vp^2 +\chi^2) =0
 \label{EOMvp} \\
 && \partial^2 \psi -2g(\chi \dot{\vp}
-\vp \dot{\chi}) -2g^2\psi (\vp^2 +\chi^2) =0
\label{EOMpsi} \\
 && \partial^2 \chi +2g(\psi \dot{\vp}
-\vp \vp') -g^2\chi (2\vp^2 -\psi^2 +\chi^2 ) =0
 \label{EOMchi} \\
 && 2g(\psi \dot{\chi} -\chi \vp') -g^2 \vp (\psi^2 +\chi^2) =0,
 \label{EOM1st}
 \eeqa
and the Lorentz gauge condition becomes
 \beq
\dot{\psi} -\vp' =0 .
 \label{LGC}
 \eeq
Here $\dot{\vp}=\partial_t \vp$, $\vp'=\partial_x \vp$, etc. Thus,
we have three second-order partial differential equations for three
unknown functions $\vp, \psi$ and $\chi$. Two first-order partial
differential equations serve as constraint equations. The energy
density in this model is given by
 \beqa
T_{00}(x,t) &=& \fr{1}{2} \left[ \left( E^a(x,t) \right)^2 +\left(
B^a(x,t) \right)^2 \right] \nonumber \\
&=& \fr{1}{2} \Bigg[ 2\left( g \chi \psi +\partial_t \vp \right)^2
+2\left( -g \vp \psi +\partial_t \chi \right)^2 +\left( \partial_t
\vp -\partial_x \psi \right)^2 +g^2 \left( 3\vp^4 +\chi^4 \right) \nonumber \\
&& +4g \vp \chi \partial_x\vp +4g \vp^2 \left( g \chi^2
-\partial_x \chi \right) +2\left( \left(
\partial_x \vp \right)^2 +\left( \partial_x \chi \right)^2 \right)
\Bigg] .
 \label{EnergyDensity}
 \eeqa

\begin{figure}[bp]
\begin{minipage}[b]{.46\linewidth}
\centering \resizebox{7cm}{7cm}{\includegraphics{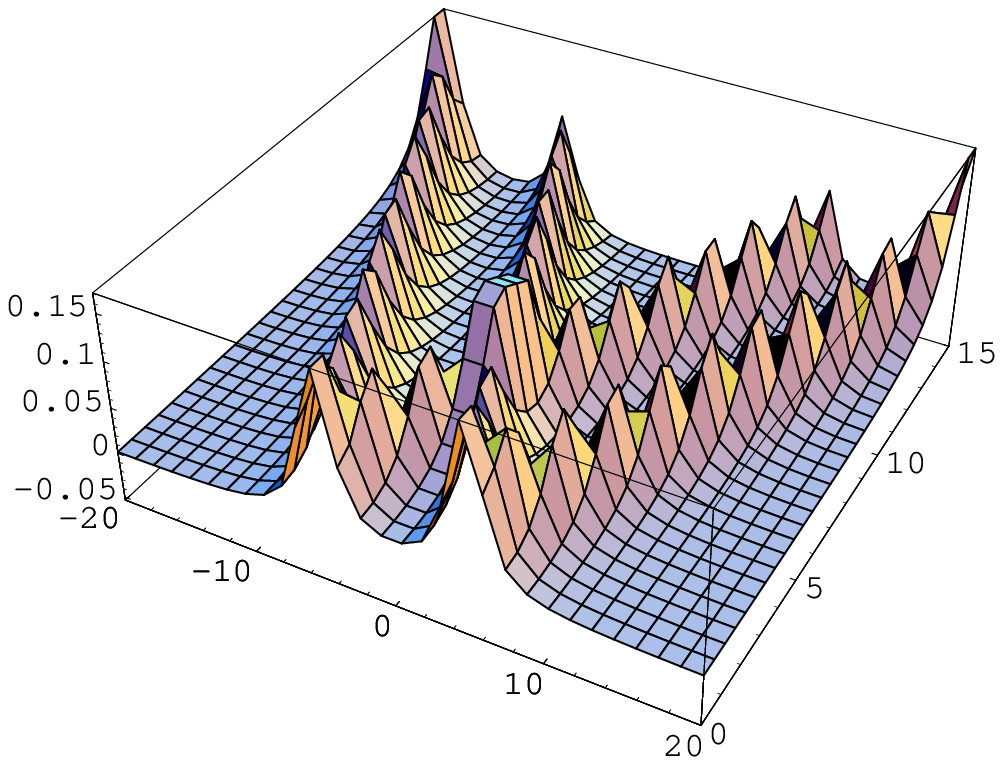}}
 \put(-135,25){$x$}
 \put(-20,60){$t$}
 \put(-198,138){$\chi(x,t)$}
 \begin{center}
(a) The evolution of the $\chi(x,t)$ field
 \end{center}
\end{minipage}\hfill
\begin{minipage}[b]{.46\linewidth}
\centering \resizebox{7cm}{7cm}{\includegraphics{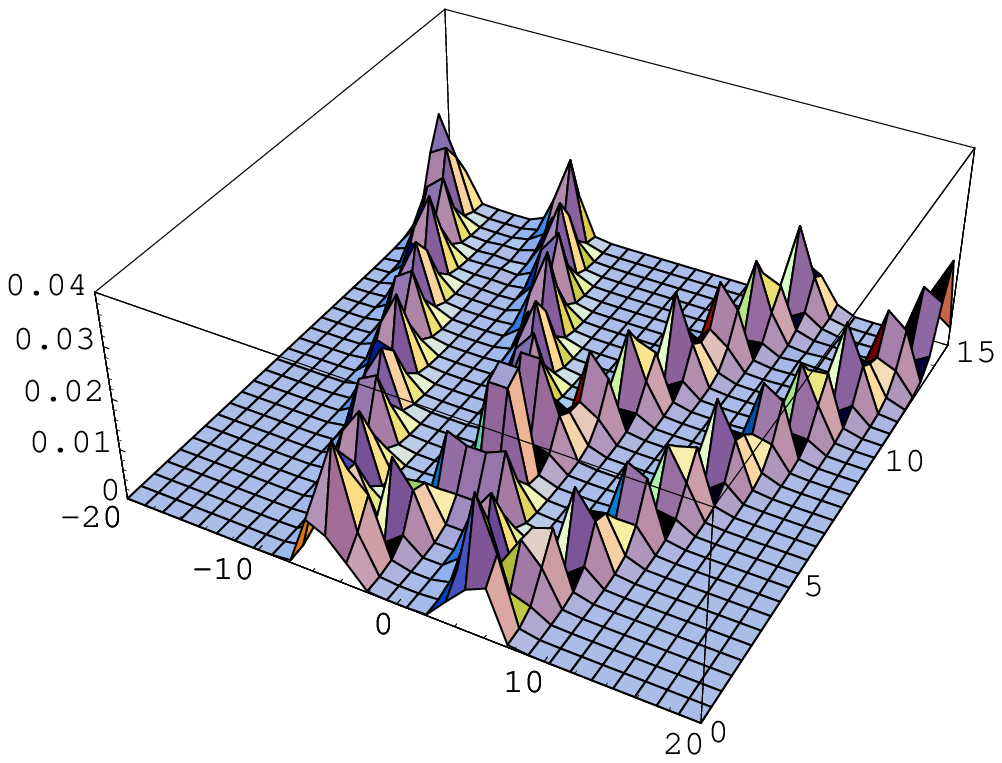}}
 \put(-135,25){$x$}
 \put(-20,60){$t$}
 \put(-198,138){$T_{00}$}
 \begin{center}
(b) Energy density $T_{00}(x,t)$
 \end{center}
\end{minipage}
\caption{The collision of bubble walls in the simplified case of
$\vp=\psi=0$ and $\chi \not= 0$}
 \label{chiDynamics}
\end{figure}

In general, it would be very hard to solve the set of equations
(\ref{EOMvp}-\ref{LGC}) above. Presumably, we have to solve the
constraint equations (\ref{EOM1st}) and (\ref{LGC}) first, and then
apply it to the dynamical equations (\ref{EOMvp}-\ref{EOMchi}) as
was done above for the case of $F$-dynamics in Eqs.~(\ref{PlaneF})
and (\ref{Errorfn}). However, we do not know how to solve the
constraint equations in general as yet.
Let us consider some special cases first. \\
$(i)$ For the case of $\chi=0$, Eq.~(\ref{EOM1st}) gives either
$\vp=0$ or $\psi=0$. If $\vp=0$, Eqs.~(\ref{EOMvp}) and
(\ref{EOMchi}) are satisfied, Eq.~(\ref{LGC}) gives
$\dot{\psi}=0$, and Eq.~(\ref{EOMpsi}) becomes equivalent to a
static potential problem in abelian theory, {\it i.e.},
$\psi''(x)=0$. If $\psi=0$, on the other hand, Eq.~(\ref{LGC})
gives $\vp'=0$, Eqs.~(\ref{EOMpsi}) and (\ref{EOMchi}) are
satisfied, and Eq.~(\ref{EOMvp}) becomes
 \beq
\ddot{\vp}(t) +2g^2 \vp^3(t) =0.
 \label{chipsizero}
 \eeq
The exact solution for this homogeneous field $\vp(t)$ is given by
the Jacobi cn function which is periodic in time. \\
$(ii)$ For the case of $\psi=0$, Eq.~(\ref{LGC}) gives $\vp'=0$ and
one finds either $\chi=0$ or $\vp=0$ from Eq.(\ref{EOM1st}). If
$\chi=0$, we have Eq.~(\ref{chipsizero}) above. If $\vp=0$, on the
other hand, Eq.~(\ref{EOMchi}) gives
 \beq
\partial^2 \chi(x,t) -g^2 \chi^3(x,t) =0.
 \label{Chi}
 \eeq
Since the constraint equations are solved already with $\vp=\psi=0$,
the remaining thing is to solve this single second order partial
differential equation only. The numerical evolution of two initial
peaks for the $\chi(x,t)$ field is shown in Fig.~\ref{chiDynamics}.
Here the initial data and boundary condition are given by
 \beq
\chi(x,t)|_{t=0}= \fr{1}{4} \left[ \fr{1}{(x-5)^2 +1}
+\fr{1}{(x+5)^2+1} \right] , \quad \partial_t\chi|_{t=0}= 0, \quad
\chi(-20,t)=\chi(20,t)
 \eeq
with $g=1$. Although a rather small peak is likely forming in the
middle, two bubble walls pass away in a short time as in the case
of linear waves. Thus, our result seems to indicate no formation
of a wall through bubble collisions in this simple model.  \\
$(iii)$ For the case of $\vp=0$, Eq.~(\ref{EOM1st}) gives either
$\psi=0$ or $\dot{\chi}=0$. If $\psi=0$, we have Eq.~(\ref{Chi})
above. If $\dot{\chi}=0$, on the other hand, Eqs.~(\ref{EOMpsi}) and
(\ref{EOMchi}) together with Eq.~(\ref{LGC}) result in
 \beq
\psi''(x) -2g^2 \psi(x) \chi^2(x) =0, \qquad \chi''(x) -g^2 \chi(x)
(\chi^2(x) -\psi^2(x) ) =0.
 \eeq
This case is a static system, which is of no interest for studying
dynamical process of bubble collisions. Thus, we find that the
non-vanishing field $\chi(x,t)$ with $\vp=\psi=0$ is the simplest
case for dynamical bubble collisions in the new Lorenzian model
(\ref{CKLmodel}), and it appears no formation of a bubble wall in
such simplified case.

\begin{figure}[tbp]
\begin{minipage}[b]{.33\linewidth}
\centering \resizebox{5cm}{5cm}{\includegraphics{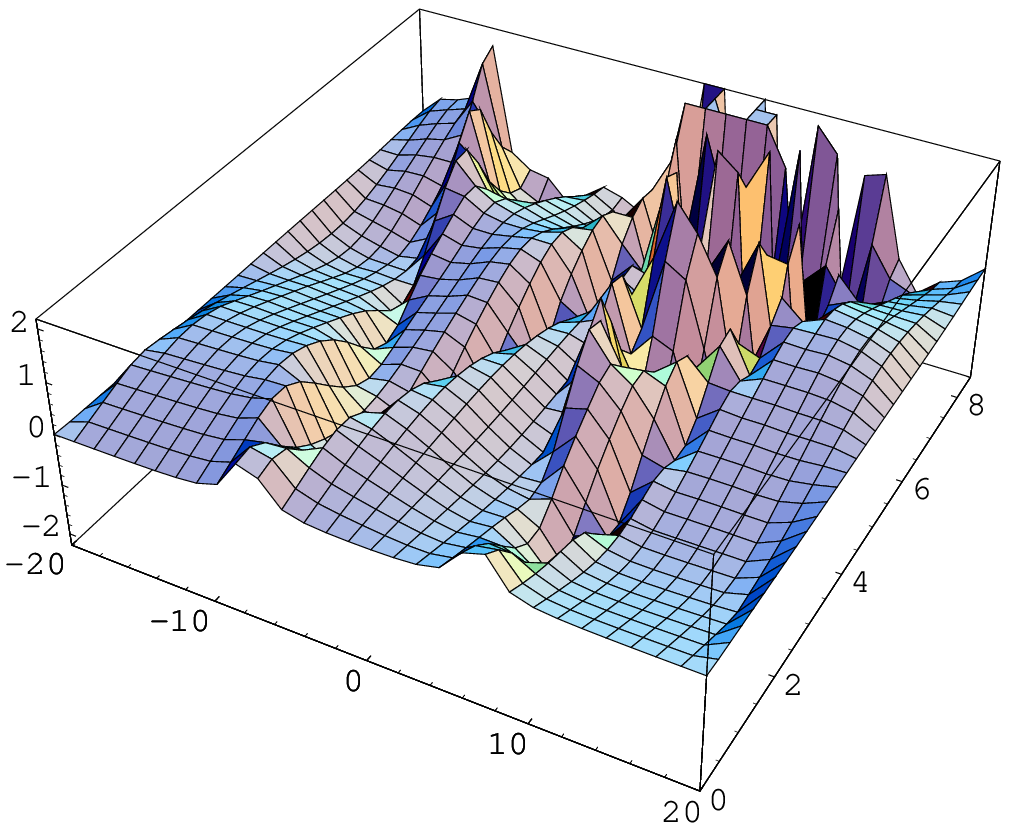}}
 \put(-100,14){$x$}
 \put(-13,42){$t$}
 \put(-141,97){$\vp$}
\end{minipage}\hfill
\begin{minipage}[b]{.33\linewidth}
\centering \resizebox{5cm}{5cm}{\includegraphics{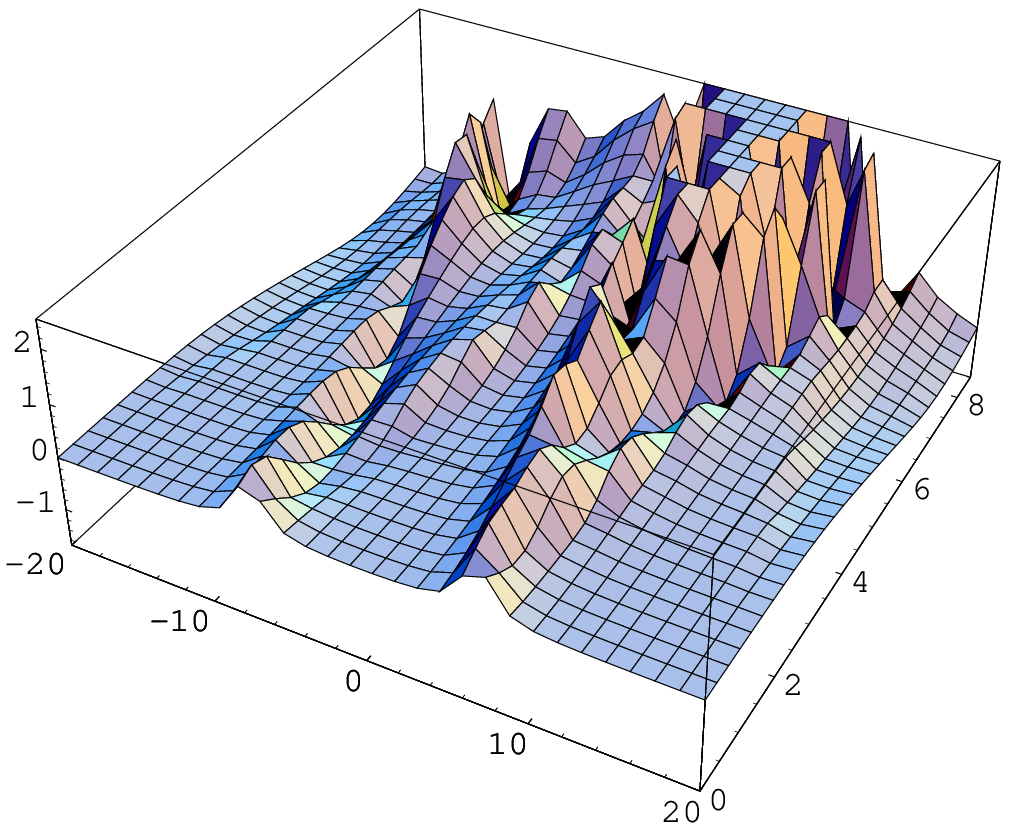}}
 \put(-100,14){$x$}
 \put(-13,42){$t$}
 \put(-141,97){$\psi$}
\end{minipage}\hfill
\begin{minipage}[b]{.33\linewidth}
\centering \resizebox{5cm}{5cm}{\includegraphics{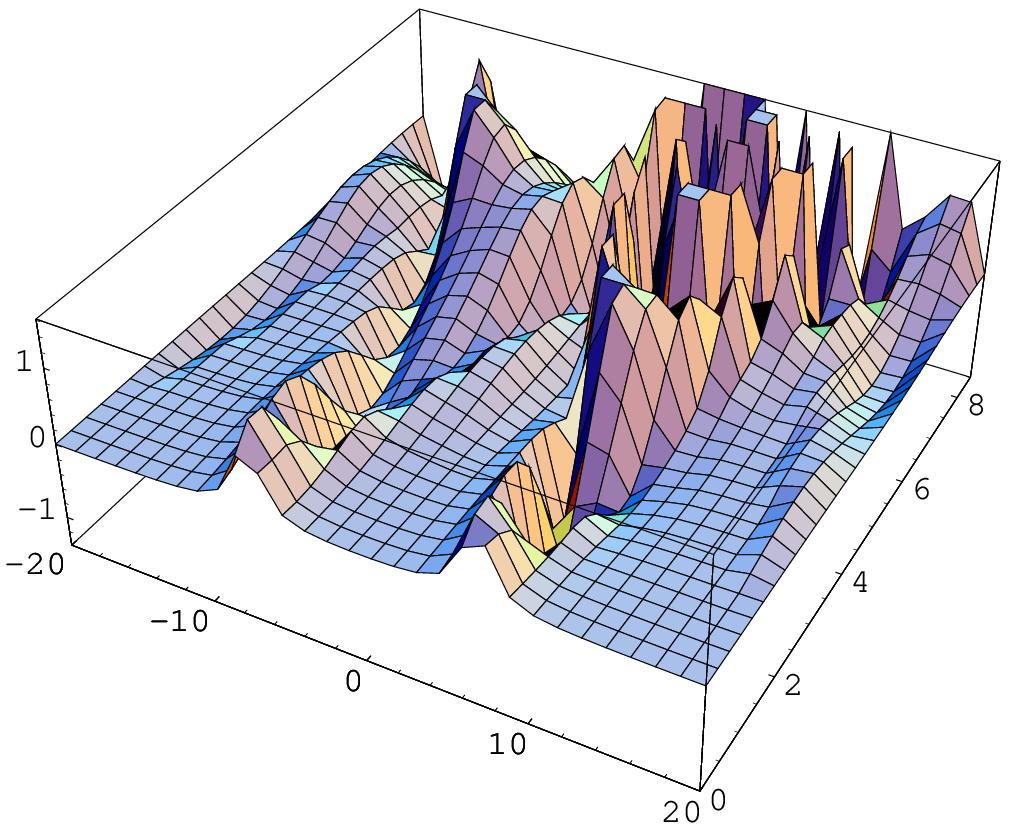}}
 \put(-100,14){$x$}
 \put(-13,42){$t$}
 \put(-141,97){$\chi$}
\end{minipage}
\begin{center}
(a) The evolution of the gluon fields  $\vp(x,t)$, $\psi(x,t)$,
and $\chi(x,t)$
\end{center}
\vspace{0.5cm}
\begin{minipage}[b]{.33\linewidth}
\centering \resizebox{5cm}{5cm}{\includegraphics{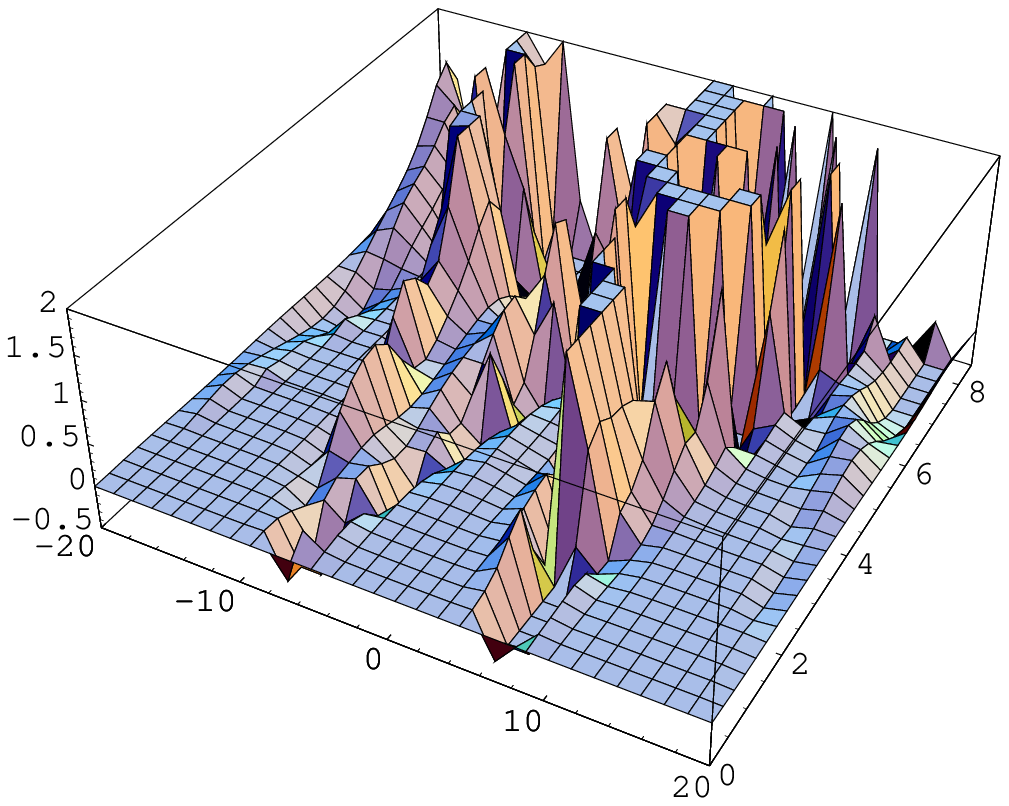}}
\begin{flushleft}
(b) Evaluation of the LHS of the constraint equation
(\ref{EOM1st})
\end{flushleft}
\end{minipage}\hfill
\begin{minipage}[b]{.33\linewidth}
\centering \resizebox{5cm}{5cm}{\includegraphics{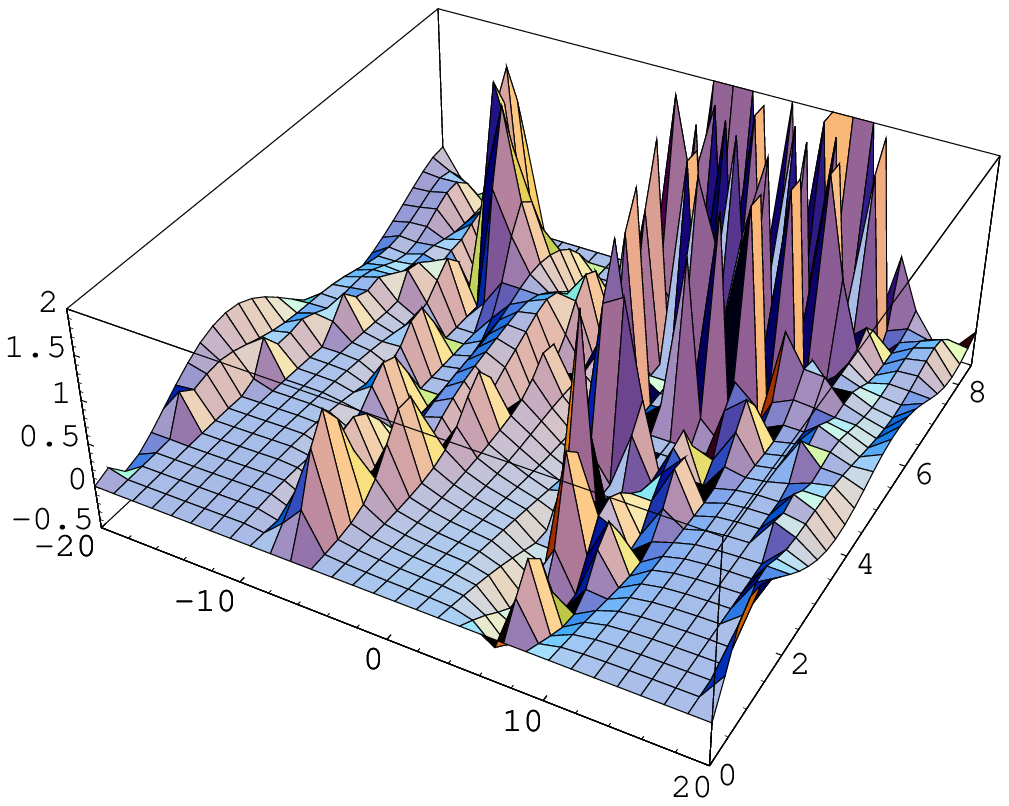}}
\begin{flushleft}
(c) Evaluation of the LHS of the constraint equation (\ref{LGC})
\end{flushleft}
\end{minipage}\hfill
\begin{minipage}[b]{.33\linewidth}
\centering \resizebox{5cm}{5cm}{\includegraphics{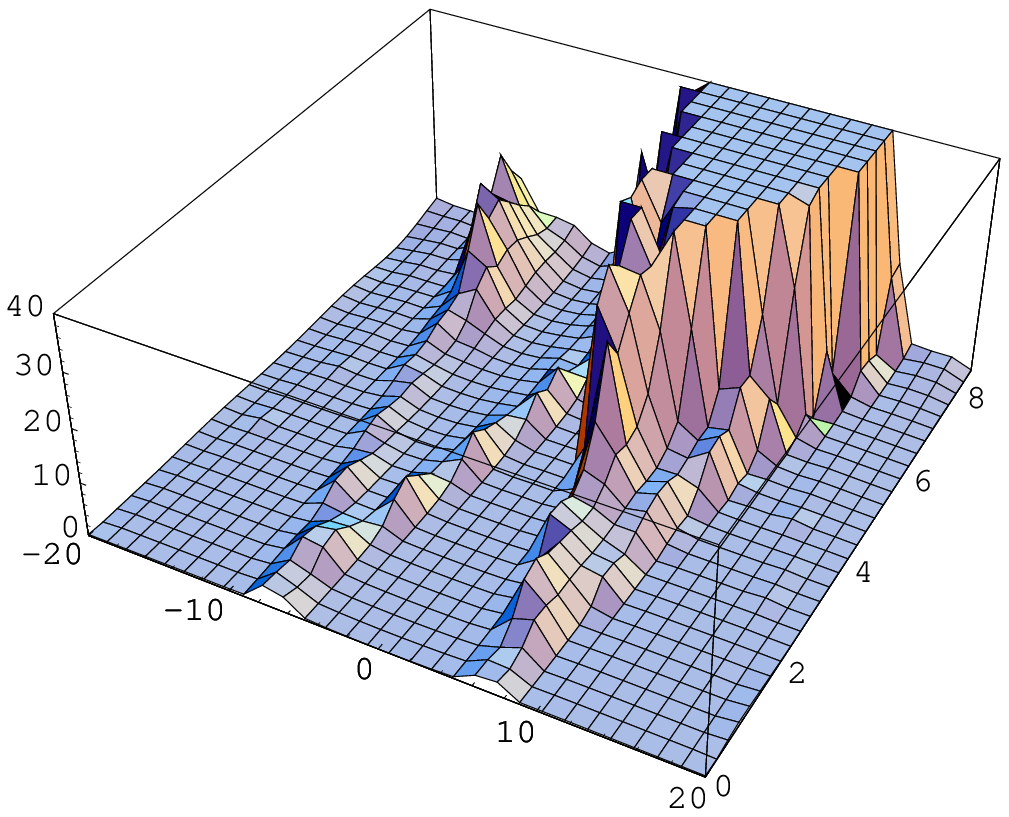}}
 \put(-100,14){$x$}
 \put(-13,42){$t$}
 \put(-141,100){$T_{00}$}
\begin{flushleft}
(d) Energy density $T_{00}(x,t)$ in Eq.~(\ref{EnergyDensity}) with
$g=1$
\end{flushleft}
\end{minipage}
\caption{The collision of bubble walls in the full model}
 \label{vpcDynamics}
\end{figure}

Now let us consider bubble collisions in the full model. Since we
are unable to solve the constraint equations in this general case
as yet, we cannot but solve dynamical equations only with
numerical method. When the coupling constant $g$ is non-vanishing,
it can be eliminated in Eqs.~(\ref{EOMvp}-\ref{LGC}) by rescaling
fields, {\it i.e.}, $\vp \rightarrow \vp/g$, $\psi \rightarrow
\psi/g$, and $\chi \rightarrow \chi/g$. Hence we set $g=1$ in the
following analysis. To solve the coupled second order partial
differential equations (\ref{EOMvp}-\ref{EOMchi}) numerically we
take initial data as
 \beqa
\mbox{} && \chi_0(x) = \chi(x,t)|_{t=0} = \fr{1}{(x-7)^2 +1}
+\fr{1}{(x+7)^2 +1}, \quad  \vp_0(x)=\psi_0(x) = \chi_0(x),
\nonumber \\
&& \dot{\vp}_0(x) = \partial_t\vp|_{t=0} = -1000 \fr{\tanh (
x/100)}{x^2 + 100}.
 \eeqa
The initial data for $\dot{\chi}_0(x)$ and $\dot{\psi}_0(x)$ are
chosen so that the first order equations (\ref{EOM1st}) and
(\ref{LGC}) are satisfied. At $x=\pm 20$ we give the periodic
boundary conditions
 \beq
\vp(-20,t) =\vp(20,t), \qquad \psi(-20,t) =\psi(20,t), \qquad
\chi(-20,t) =\chi(20,t).
 \eeq
Our numerical results for this set of initial data are shown in
Fig.~\ref{vpcDynamics}. The evolution of the gluon fields $\vp$,
$\psi$ and $\chi$ can be seen in Fig.~\ref{vpcDynamics} (a) in the
range of time: $t=0 \sim 8.51$, and the change of the
corresponding energy density is shown in (d). In
Fig.~\ref{vpcDynamics} (d) a wall seems to appear at around $t
\sim 4.5$. Since the constraint equations (\ref{EOM1st}) and
(\ref{LGC}) are almost fulfilled at least until around $t \sim
4.5$ as can be checked in Fig.~\ref{vpcDynamics} (b) and (c), this
may indicate a formation of bubble wall indeed.

\begin{figure}[tbp]
\begin{minipage}[b]{.33\linewidth}
\centering \resizebox{5cm}{5cm}{\includegraphics{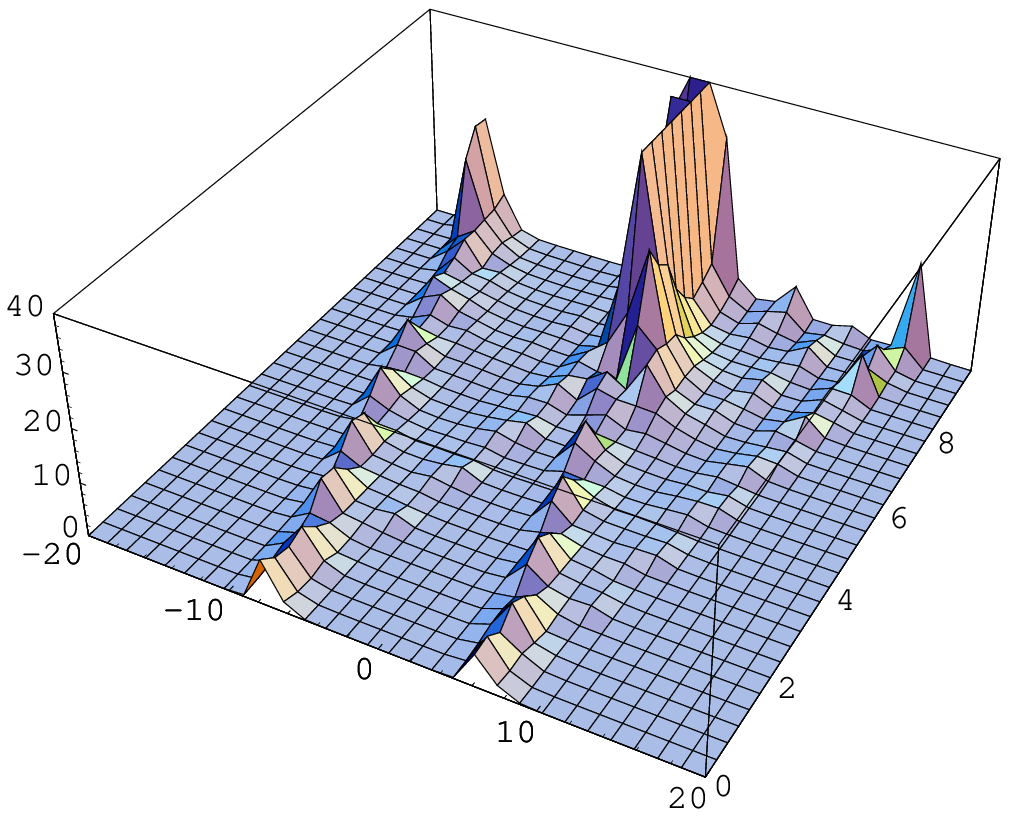}}
 \put(-100,14){$x$}
 \put(-13,42){$t$}
 \put(-141,100){$T_{00}$}
\begin{center}
(a) $\vp_0=\psi_0=0.3\chi_0$
\end{center}
\end{minipage}\hfill
\begin{minipage}[b]{.33\linewidth}
\centering
\resizebox{5cm}{5cm}{\includegraphics{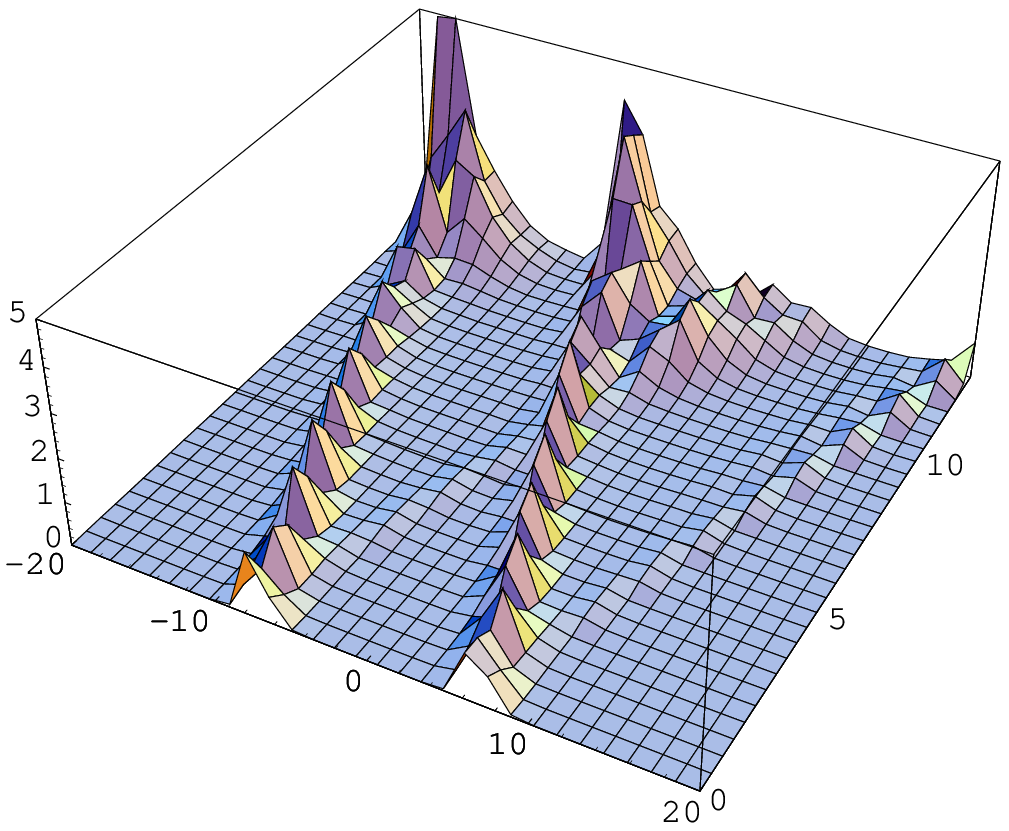}}
 \put(-100,14){$x$}
 \put(-13,42){$t$}
 \put(-141,100){$T_{00}$}
\begin{center}
(b) $\vp_0=\psi_0=0.05\chi_0$
\end{center}
\end{minipage}\hfill
\begin{minipage}[b]{.33\linewidth}
\centering \resizebox{5cm}{5cm}{\includegraphics{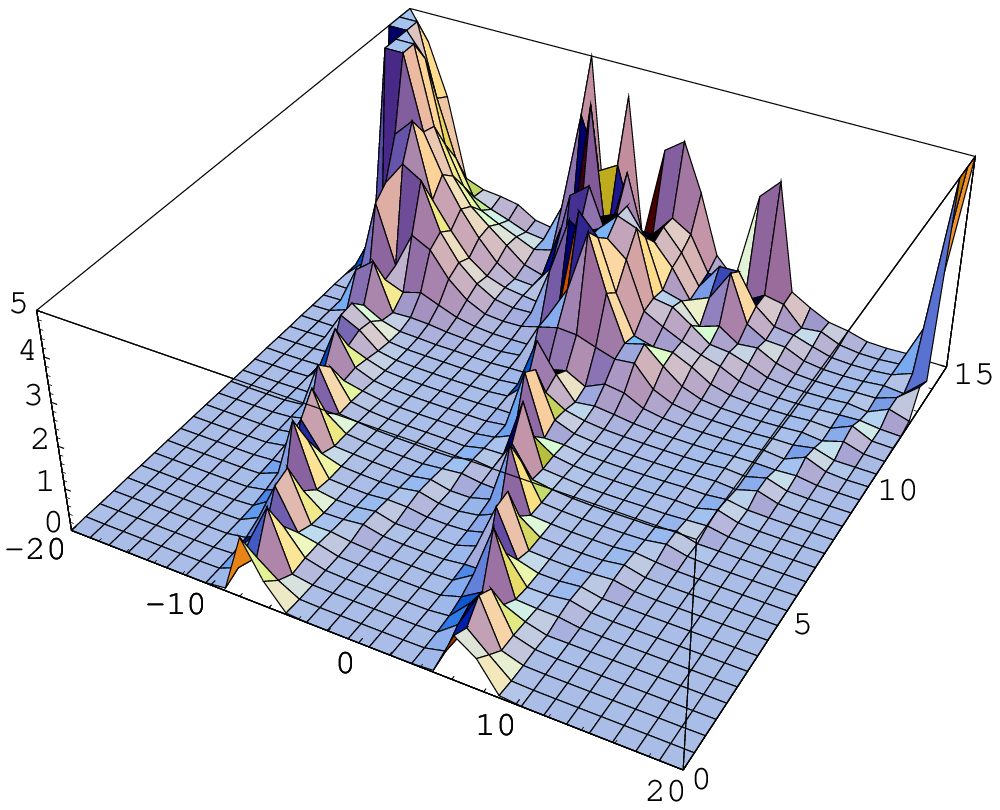}}
 \put(-100,14){$x$}
 \put(-13,42){$t$}
 \put(-141,100){$T_{00}$}
\begin{center}
(c) $\vp_0=\psi_0=0.01\chi_0$
\end{center}
\end{minipage}
\caption{The energy density for various initial values of $\vp$
and $\psi$. The formation of bubble wall appears at around $t \sim
7.5,\, 10.5$, and $11.0$, respectively.}
 \label{BubbleCollision}
\end{figure}

We also have considered some other sets of initial data.
Fig.~\ref{BubbleCollision} shows our numerical results for
$\vp_0=\psi_0 = 0.3 \chi_0$, $0.05 \chi_0$, and $0.01 \chi_0$ with
$\dot{\vp}_0 = -100 \tanh ( x/100)/(x^2 + 100)$. Formation of
bubble wall occurs for all three cases as well although the moment
of bubble wall formation differs depending on the value of initial
data. As in the previous case, the first-order constraint
equations are almost fulfilled until around the time from which
bubble wall starts to form. Note that, in a longer time, the
evolution of two initial peaks for small $\vp_0$ and $\psi_0$ is
very much different from that for the case of vanishing $\vp$ and
$\psi$ in Eq.~(\ref{Chi}) even if they are similar in early time.
It implies that the presence of all non-vanishing three fields
$\vp$, $\psi$ and $\chi$ in our model is necessary for the
formation of bubble wall during collisions.


\section{Conclusions and Discussions}

In this work, we have considered the collision of gluonic bubbles
in the context of the instanton-inspired model of QCD phase
transition bubbles discussed in
Refs.~\cite{Kisslinger:2002py,Kisslinger:2002mu,Johnson:2003ti}.
In particular, it has been investigated the possibility of bubble
wall formation during pure gluonic bubble collisions in plane wave
approximation.

Since we are dealing with a system of coupled nonlinear equations
of the color dynamics, there exist both dynamical equations and
constraint equations in this color dynamics. Correct approach to
solve these kinds of system is to solve the constraint equations
first and then apply to the dynamical equations. In the case of
$F$-dynamics in Eqs.~(\ref{PlaneF}) and (\ref{Errorfn}), we can
find even an exact solution in this way. Our $F$-dynamics case
corresponds to the holonomic constrained system. For other cases,
the problems are much harder because the constraints are not
easily removed. At present, we cannot but solve these problems
with numerical method. However, we checked that our numerical
results satisfy the first-order constraint equations, up to some
small error, till the appearance of bubble wall formation.

With this correct approach described above we reanalyzed the
$F$-dynamics in Eqs.~(\ref{PlaneF}) and (\ref{Errorfn}), and found
no indication of the bubble wall formation in colliding processes.
Moreover, it is pointed out that the instanton-inspired model
studied in
Refs.~\cite{Kisslinger:2002py,Kisslinger:2002mu,Johnson:2003ti}
({\it i.e.}, Eq.~(\ref{RelationshipLE}) with $W_2=W_3=0$) should
be corrected due to the presence of imaginary gluon fields, since
it leads to the violation of causality. Therefore, we reconsidered
the process of bubble collisions in a new slightly modified
Lorentzian model (\ref{CKLmodel}) where we have three independent
real gluon fields. For the case of $\chi$-dynamics with
$\vp=\psi=0$ in Eq.~(\ref{Chi}), we did not find any evidence of
the bubble wall formation. For more complicated cases in
Eqs.~(\ref{EOMvp})-(\ref{LGC}), however, we see some indication of
forming the bubble wall. It is likely that the presence of all
non-vanishing three gluon fields is necessary to have the
formation of bubble wall in collisions.

Nevertheless, we stress that much more careful analyses are
necessary to confirm our results. Although the bubble wall
formation occurs in the colliding region of two peaks, the
location is not exactly the colliding point except for the case of
Fig.~\ref{BubbleCollision} (a). As can be seen in
Fig.~\ref{vpcDynamics} (b) and (c), the first-order equations are
not satisfied well beyond the time of bubble wall appearance.
Consequently, the numerical results beyond this time may not be
trustworthy in Figs.~\ref{vpcDynamics} and \ref{BubbleCollision}.
The unexpected appearance of peaks at the left end in late time
evolution in Figs.~\ref{vpcDynamics} and \ref{BubbleCollision} may
be related to this increasing error as time goes by. It will be of
great interest to see how the bubble wall formed actually evolves
afterwards.

Investigations on solving the constraint equations for the
complicated cases are in progress. The idea of applying canonical
transformations to the color-dynamic constrained system as in the
case of $F$-dynamics may deserve further consideration.

\section*{Acknowledgments}

We thank Ho-Meoyng Choi for many useful discussions. One of us (J. Lee)
also would like to acknowledge helpful discussions with Youngduk Han.
This work was supported in part by a grant from the U.S.Department
of Energy (DE-FG02-96ER 40947). The National Energy Research Scientific
Computing Center(NERSC) is also acknowledged for the computing time.

\section*{APPENDIX: More Details of Error Estimates}

Since the analysis based on Eq.~(\ref{Fform}) for bubble
collisions is severely constrained both in the number of degrees
of freedom and the form of the function $F(x,t)$, one may consider
an extension to the study with the larger degrees of freedom using
$W_0$ and $W_1$. Note that the set of equations (\ref{eom1}) and
(\ref{gcondx}) for $W_0$ and $W_1$ consists of two coupled
second-order partial differential equations with one first-order
equation in time. As explained in the case of $F$-dynamics in
Eqs.~(\ref{PlaneF}) and (\ref{Errorfn}), the solution for
Eq.~(\ref{eom1}) should satisfy the constraint equation
(\ref{gcondx}) as well. In general, not all solutions for
Eq.~(\ref{eom1}) fulfill Eq.~(\ref{gcondx}). Since it is not easy
to solve the constraint equation for this case, in contrast to the
case of $F$-dynamics, we cannot but solve the dynamical equations
(\ref{eom1}) numerically for certain initial data.

Since the evolution equations in (\ref{eom1}) are second-order,
one can take any initial data set of $W_0(x,0)$, $W_1(x,0)$,
$\partial_tW_0 =\dot{W}_0(x,0)$ and $\partial_tW_1
=\dot{W}_1(x,0)$ at $t=0$ subject to the first-order equation
(\ref{gcondx}) evaluated at $t=0$ being satisfied ({\it i.e.},
$\dot{W}_1(x,0)=\partial_xW_0(x,0)$). The case of $F$-dynamics in
Eq.~(\ref{InitialDF}) corresponds to the special case of
$W_1(x,0)=\fr{2}{g}xF(x,0)$, $W_0(x,0)=\dot{W}_1(x,0)=0$, and
$\dot{W}_0(x,0)=-\fr{2}{g}F(x,0)$. In Ref.~\cite{Johnson:2003ti}
the authors consider a somewhat restricted set of initial data
such that $W_0 (x,0)=c$, $W_1 (x,0)=w(x)$, $\dot{W}_0(x,0)=0$, and
$\dot{W}_1 (x,0) =0$.~\footnote{The description for the initial
data in Ref.~\cite{Johnson:2003ti} is somewhat misleading as we
clarify it in this work.} This initial data set satisfies the
gauge condition Eq.~(\ref{gcondx}), and the equations of motion
evaluated at $t=0$ are written as
\begin{eqnarray}
 \label{weq1}
\partial_x w(x)= g (w^2-c^2)  % +\fr{1}{c}w \dot{W}_1(x,0)
+ \frac{1}{2gc} \ddot{W}_0(x,0), \\
 \label{weq2}
\partial^2_x w(x)= 2g^2 w(w^2-c^2)  % +2gw\dot{W}_0(x,0)
+\ddot{W}_1(x,0),
\end{eqnarray}
through which the values of $\ddot{W}_0(x,t)$ and
$\ddot{W}_1(x,t)$ at $t=0$ can be determined. However, the authors
in Ref.~\cite{Johnson:2003ti} restricted the initial data set
further somehow into the case of
$\ddot{W}_0(x,0)=\ddot{W}_1(x,0)=0$. For such case the function
$w(x)$ is determined as
 \begin{eqnarray}
 \label{wsol1}
w(x)= -c \tanh [c g (x-x_0 ) ].
 \end{eqnarray}
Here $x_0$ is an integration constant.

As mentioned above, however, one can take in general any function
$w(x)$ as an initial profile subject to some unknown restriction
due to the presence of the unsolved constraint equation. Authors
in Ref.~\cite{Johnson:2003ti} used a symmetric ansatz for an
initial profile of $w(x)$ given by
\begin{eqnarray}
 \label{wsol2}
w(x)= -c \left\{ 1/2 +\tanh \left[ c g (x -x_0 )\right] \right\}
-c \left\{ 1/2 -\tanh \left[ c g(x +x_0 ) \right] \right\}
\end{eqnarray}
with $c=4$, $g=1$, and $x_0=3$. They considered the symmetric
boundary condition, $W_i (-10, t) = W_i (10, t)$, and vanishing
initial velocities, $\partial_t W_i (x,0)=0$, for $i= 0, \, 1$. In
this plane wave model, the energy-momentum tensor density
$T_{00}(x,t)$ is given by
 \begin{eqnarray}
 \label{pemt}
T_{00}(x,t) &=& 3 (\partial_t W_0)^2 + (\partial_x W_1)^2 -2
\partial_t W_0 \partial_x W_1 -(\partial_t W_1)^2 -(\partial_x W_0)^2
\nonumber \\
&& + 2g \left[ 2 W_1^2 \partial_t W_0 -W_0 W_1(\partial_x W_0
+\partial_t W_1) \right] + 2g^2 W_1^2 (W_1^2 - W_0^2) + {\cal
L}^{\rm glue}
\end{eqnarray}
in terms of the fields $W_0(x,t)$ and $W_1(x,t)$. Here ${\cal
L}^{\rm glue}$ represents the gluonic Lagrangian density given by
 \begin{eqnarray}
 \label{plag}
{\cal L}^{\rm glue} &=& -\fr{1}{2} \bigg\{ 3(\partial_t W_0)^2
+3(\partial_x W_1)^2 -2(\partial_t W_1)^2 -2(\partial_x W_0)^2 -2
\partial_t W_0 \partial_x W_1 \nonumber\\
&& +4g \left[ W_1^2 \partial_t W_0 +W_0^2 \partial_x W_1 -W_0 W_1
(\partial_t W_1 +\partial_x W_0) \right] +3g^2 (W_1^2 -W_0^2)^2
\bigg\} .
\end{eqnarray}
Note that the Lagrangian density in Ref.~\cite{Johnson:2003ti} has
the opposite sign of it with the absence of the fifth term above.

\begin{figure}[tbp]
\begin{minipage}[b]{.33\linewidth}
\centering
\resizebox{5cm}{5cm}{\includegraphics{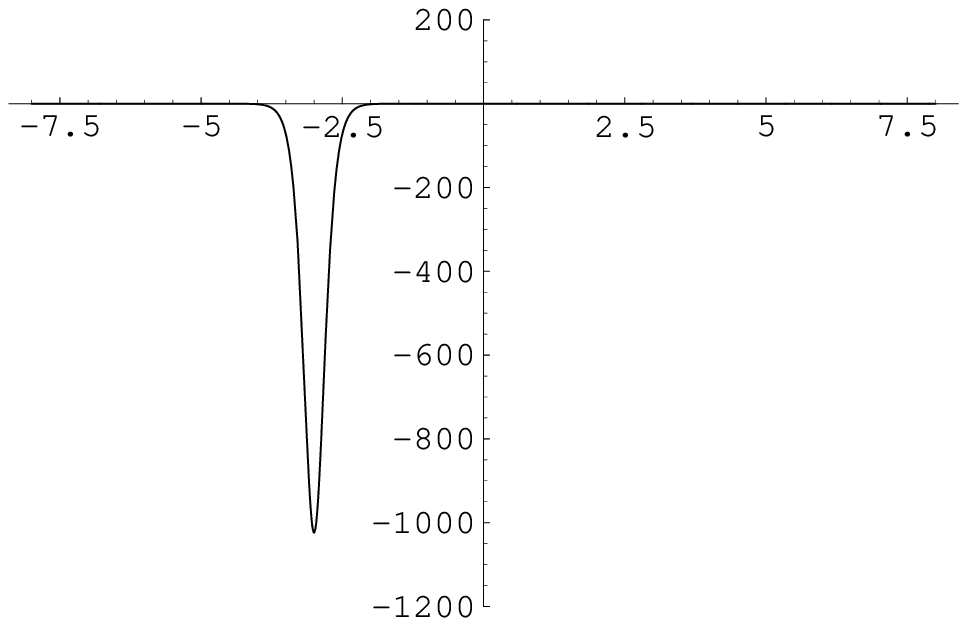}}
 \put(-4,117){$x$}
 \begin{center}
(a) $T_{00}(x,0)$
 \end{center}
\end{minipage}\hfill
\begin{minipage}[b]{.33\linewidth}
\centering
\resizebox{5.3cm}{5.3cm}{\includegraphics{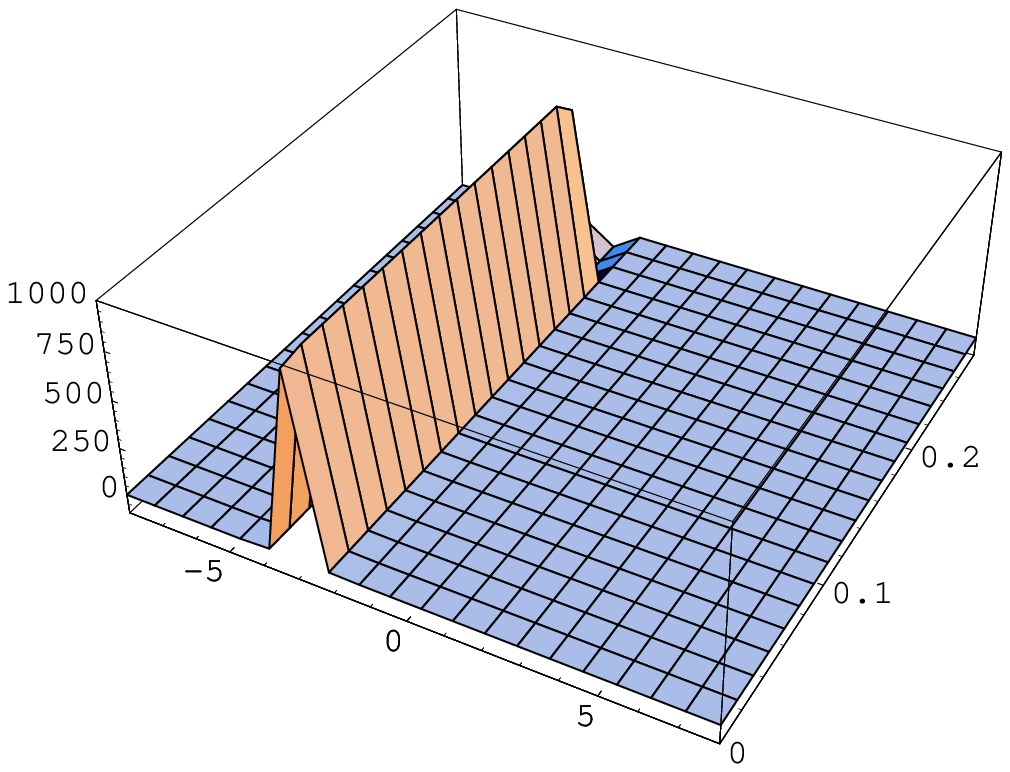}}
 \put(-100,15){$x$}
 \put(-14,46){$t$}
 \put(-152,105){$-T_{00}$}
 \begin{center}
(b) $-T_{00}(x,t)$
 \end{center}
\end{minipage}\hfill
\begin{minipage}[b]{.33\linewidth}
\centering
\resizebox{5.3cm}{5.3cm}{\includegraphics{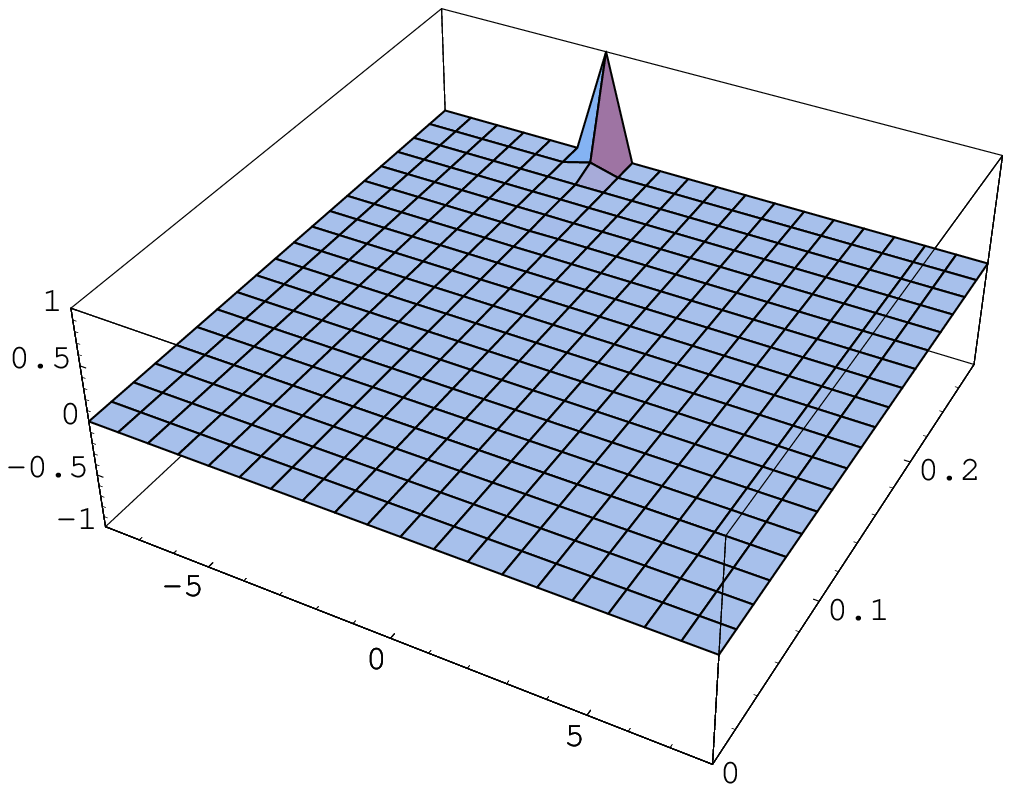}}
 \put(-100,15){$x$}
 \put(-14,46){$t$}
 \begin{center}
(c) $\partial_x W_0 -\partial_t W_1$
 \end{center}
\end{minipage}
\caption{A numerical solution up to $t=0.28$ in the dynamics of
$W_0$ and $W_1$}
 \label{figW0W1028}
\end{figure}

\begin{figure}[tbp]
\begin{minipage}[b]{.33\linewidth}
\centering
\resizebox{5.3cm}{5.3cm}{\includegraphics{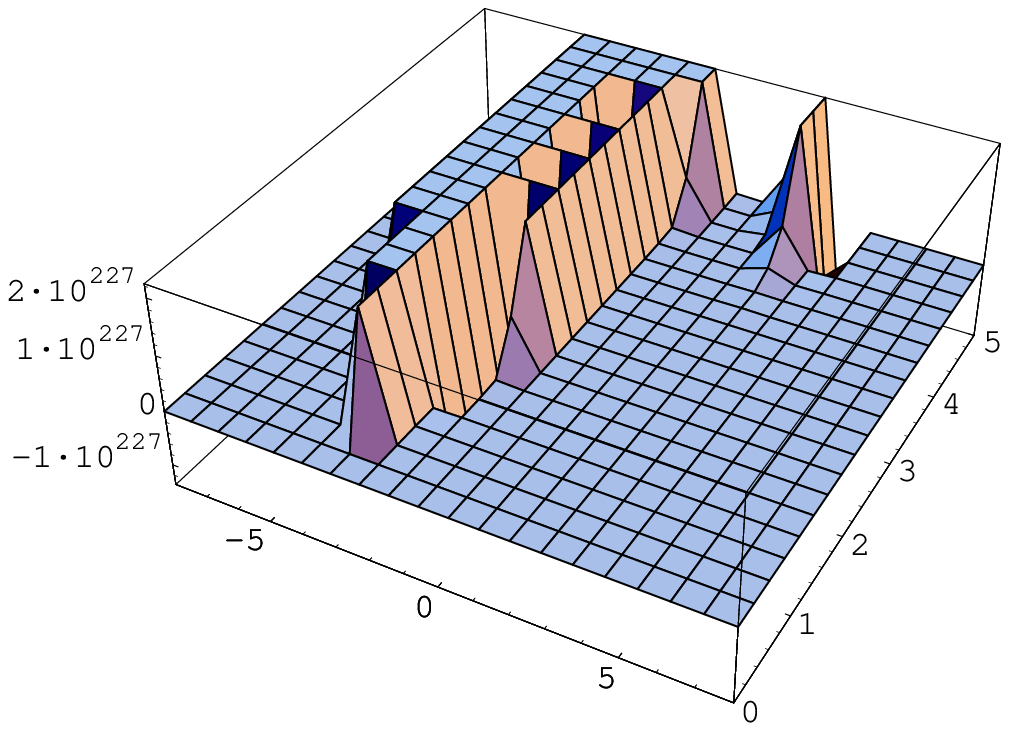}}
 \put(-99,16){$x$}
 \put(-12,45){$t$}
 \put(-152,105){$-T_{00}$}
 \begin{center}
(a) $-T_{00}(x,t)$
 \end{center}
\end{minipage}\hfill
\begin{minipage}[b]{.33\linewidth}
\centering
\resizebox{5.3cm}{5.3cm}{\includegraphics{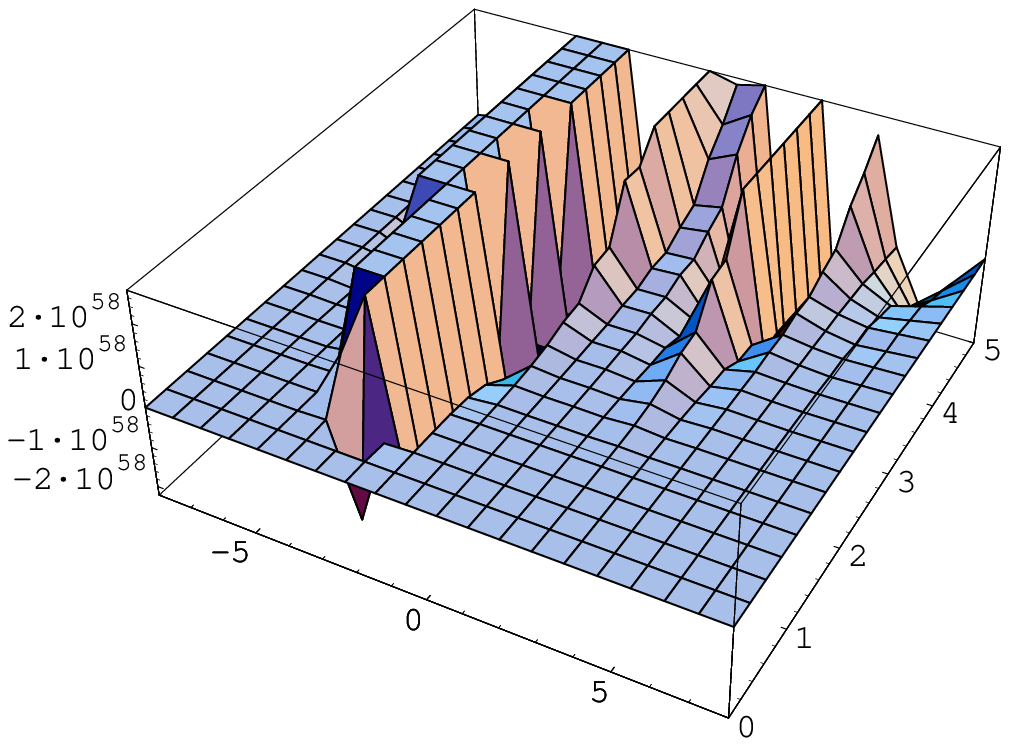}}
 \put(-99,16){$x$}
 \put(-12,45){$t$}
 \begin{center}
(b) $\partial_x W_0 -\partial_t W_1$
 \end{center}
\end{minipage}\hfill
\begin{minipage}[b]{.33\linewidth}
\centering
\resizebox{5cm}{5cm}{\includegraphics{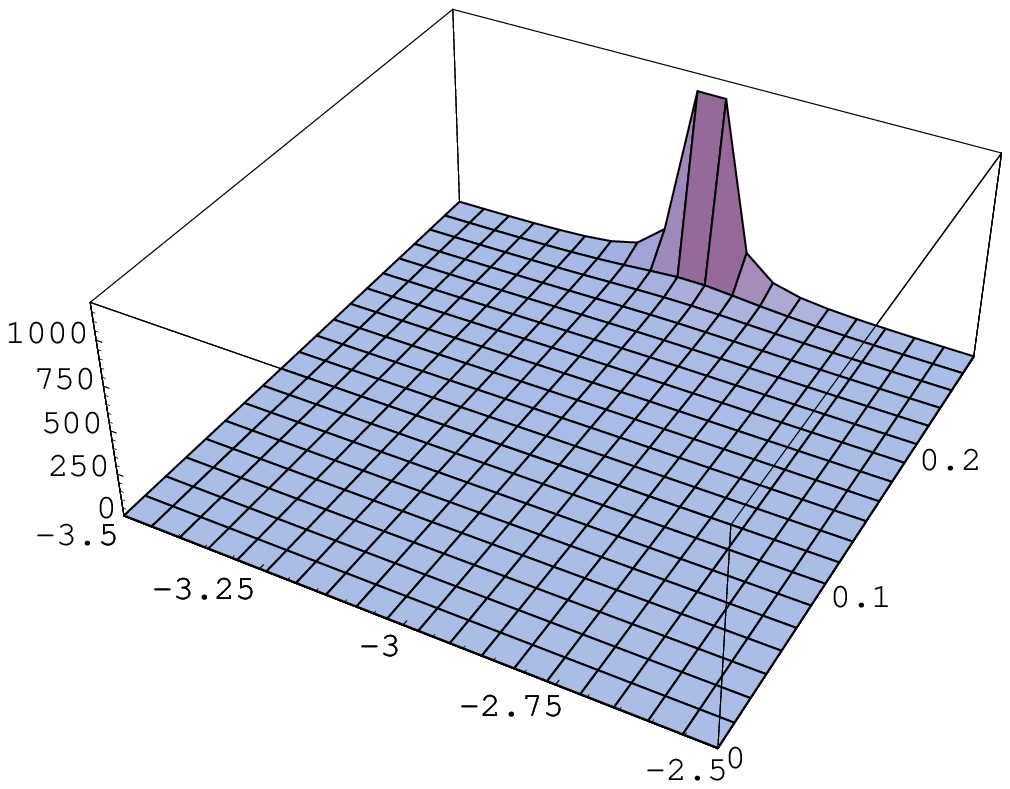}}
 \put(-99,16){$x$}
 \put(-12,45){$t$}
 \put(-142,98){$W_0$}
 \begin{center}
(c) $W_0(x,t)$
 \end{center}
\end{minipage}\hfill
\caption{The extrapolated solution up to $t=5$ in the dynamics of
$W_0$ and $W_1$}
 \label{figW0W15}
\end{figure}

As can be seen in Fig.~\ref{figW0W1028} (a), we point out that the
energy profile corresponding to the initial data given above is
not in the form of two lumps that authors in
Ref.~\cite{Johnson:2003ti} expected. The numerical solution
evolved up to $t=0.28$ is shown in Fig.~\ref{figW0W1028} (b). One
can see in Fig.~\ref{figW0W1028} (c) that the constraint equation
$\partial_x W_0 -\partial_t W_1=0$ is satisfied well up to $t
\simeq 0.28$ since $|\partial_x W_0 -\partial_t W_1|$ is very
small in comparison with $|\partial_x W_0|, \, |\partial_tW_1|$
whose numerical values turn out to be the order of $\sim 1$. This
numerical calculation was able to be performed up to $t=0.28$ by
using the NDSolve program in MATHEMATICA. However, if we request
plotting the result further beyond $t=0.28$, MATHEMATICA gives it
by performing extrapolations which are illustrated in
Fig.~\ref{figW0W15}.

One can see that this extrapolated energy density changes
drastically around $t=0.28$ from the order of $10^3$ to the order
of $10^{227}$, indicating a singular behavior of solution. One
might think that the peak shown in Fig.~\ref{figW0W15} (a) is an
indication of the bubble wall formation as discussed in
Ref.~\cite{Johnson:2003ti}. However, it is hard to think that an
initially single peak develops a bubble wall. The consideration of
other type of initial data giving an energy profile of two peaks
would be necessary. Fig.~\ref{figW0W15} (b) also shows that the
constraint equation is not satisfied well for this extrapolated
solution since $|\partial_x W_0|, \, |\partial_tW_1|$ becomes the
order of $10^{58}$ which is almost same as the order of
$|\partial_x W_0 -\partial_t W_1|$. For these reasons, therefore,
we think that more careful analysis is needed on the bubble
collision in the dynamics of $W_0$ and $W_1$.

\end{document}